\documentclass[prb,showpacs,preprintnumbers,amsmath,amssymb,superscriptaddress]{revtex4}
\usepackage{epsfig}
\usepackage{graphicx}
\usepackage{dcolumn}
\usepackage{bm}

\begin{document}
\title{Magnetic state of plutonium ion in metallic Pu and its compounds}
\author{A.O.~Shorikov}
\affiliation{Institute of Metal Physics, Russian Academy of Sciences-Ural Division,
620219 Yekaterinburg GSP-170, Russia}
\author{A.V.~Lukoyanov}
\affiliation{Institute of Metal Physics, Russian Academy of Sciences-Ural Division,
620219 Yekaterinburg GSP-170, Russia}
\affiliation{Ural State Technical University-UPI,
620002 Yekaterinburg, Russia}
\author{M.A.~Korotin}
\affiliation{Institute of Metal Physics, Russian Academy of
Sciences-Ural Division, 620219 Yekaterinburg GSP-170, Russia}
\author{V.I.~Anisimov}
\affiliation{Institute of Metal Physics, Russian Academy of Sciences-Ural Division,
620219 Yekaterinburg GSP-170, Russia}

\date{\today}
\begin{abstract}
By LDA+U method with spin-orbit coupling (LDA+U+SO) the magnetic
state and electronic structure have been investigated for
plutonium in $\delta$ and $\alpha$ phases and for Pu compounds:
PuN, PuCoGa$_5$, PuRh$_2$, PuSi$_2$, PuTe, and PuSb. For metallic
plutonium in both phases in agreement with experiment a
nonmagnetic ground state was found with Pu ions in $f^6$
configuration with zero values of spin, orbital, and total
moments. This result is determined by a strong spin-orbit coupling
in 5$f$ shell that gives in LDA calculation a pronounced splitting
of 5$f$ states on $f^{5/2}$ and $f^{7/2}$ subbands. A Fermi level
is in a pseudogap between them, so that $f^{5/2}$ subshell is
already nearly completely filled with six electrons before Coulomb
correlation effects were taken into account. The competition
between spin-orbit coupling and exchange (Hund) interaction
(favoring magnetic ground state) in 5$f$ shell is so delicately
balanced, that a small increase (less than 15\%) of exchange
interaction parameter value from $J_H$ = 0.48~eV obtained in
constrain LDA calculation would result in a magnetic ground state
with nonzero spin and orbital moment values. For Pu compounds
investigated in the present work, predominantly $f^6$
configuration with nonzero magnetic moments was found in
PuCoGa$_5$, PuSi$_2$, and PuTe, while PuN, PuRh$_2$, and PuSb have
$f^5$ configuration with sizeable magnetic moment values. Whereas
pure $jj$ coupling scheme was found to be valid for metallic
plutonium, intermediate coupling scheme is needed to describe
5$f$ shell in Pu compounds. The results of our calculations show
that exchange interaction term in the Hamiltonian must be treated 
in a general matrix form for Pu and its compounds.
\end{abstract}


\maketitle
\section{Introduction}
In actinide elements Coulomb interaction of 5$f$ electrons is of
the same order as a band width and spin-orbit coupling. Their interplay
results in a complicated physics of 5$f$ compounds~\cite{Freeman74}
where both theorists and experimentalists still
have attractive plenty of work. Among other actinides, plutonium
element seams to be most intriguing,~\cite{HeckerPMS,Lander03}
despite numerous works and 60 years passed since the Pu
discovery.~\cite{LosAlamos,MRS}

Electronic properties of plutonium show an exceptional example of
the system with 5$f$ electrons on the edge between localization
and itinerancy.~\cite{Lander03,Albers01} Experimental
work\cite{Moore03} gave $jj$ coupling for Pu 5$f$ electrons or an
intermediate coupling close to $jj$ type.~\cite{Laan04}
Theoretical model for intermediate coupling scheme was suggested 
for PuSb by Cooper {\it et al.}~\cite{Cooper85}

Many experiments have been carried out to shed light on plutonium
electronic structure. For $\delta$-plutonium, most investigated of
all metallic Pu phases, photoemission spectroscopy (PES) revealed
special features of valence band~\cite{PESa,PESb,Naegele86} in
particular a sharp peak at the Fermi level. X-ray, high-resolution
ultraviolet,~\cite{Havela02} and resonant
photoemission\cite{Tobin03} spectroscopy measurements accompanied
with Pu 4$f$ core-level spectra show more localized character of
5$f$ electrons in $\delta$-Pu phase comparing with $\alpha$ phase.
Sharp peak at the Fermi level is indicative of strong
many-particle nature and heavy-fermion behavior of 5$f$
electrons. The electron mass enhancement is usually characterized
by the inferred Sommerfeld coefficient $\gamma$ = 50 mJ mol$^{-1}$
K$^{-2}$ for $\delta$-Pu\cite{Wick80} versus $\gamma$ = 17 mJ
mol$^{-1}$ K$^{-2}$ for $\alpha$-Pu.~\cite{Lashley03} Phonon
dispersions and elastic moduli of $\delta$-Pu showed a number of
anomalies,~\cite{Wong03,Ledbetter76} including softening of the
transverse [111] modes,~\cite{McQueeney04} suggesting unusual
electronic structure and phonon-electron interaction for 5$f$
electrons in plutonium (modern calculation results\cite{Dai03}
qualitatively predicted experimental data). No evidence for
ordered or disordered magnetic moments in plutonium in both
$\delta$ and $\alpha$ phases was found in
experiments.~\cite{Lashleycondmat,HeckerPMS,LosAlamos}

{\it Ab-initio} calculations of the electronic structure for
plutonium have been done by many authors.~\cite{Richter01} Standard
local density approximation (LDA)\cite{Ek93} as well as
generalized gradient approximation (GGA)~\cite{Soderlind97} have
been implemented. Full-potential linearized augmented plain-wave method
(FLAPW) with GGA-correction
and Gaussian-type orbitals-fitting function (LCGTO-FF) methods
were used to reproduce zero-pressure experimental volumes and bulk
module of Pu in the row of all actinides.~\cite{Jones00} The
obtained results show better agreement with experiment than, for
example, LMTO-GGA method\cite{Soderlind94} but fails handling with
Pu. Even with proper second variational approach to $6p$
states,~\cite{Fernando00} experimental volume of $\delta$-Pu is
still unattainable.~\cite{Nordstrom01} Another limitation of all
these methods, as it was realized later, is inability to reproduce
correctly electronic structure of plutonium due to neglect of
Coulomb correlation effects.

To improve LDA approximation LDA+U method\cite{LDA+U} was used,
which explicitly takes into account strong Coulomb repulsion between
5$f$ electrons by adding Hubbard-like term to the LDA Hamiltonian.
The calculation of electronic structure for fcc-Pu by LDA+U
method with GGA correction was performed in Ref.~\onlinecite{Savrasov00}. 
Authors showed that for the standard set of Slater
integrals\cite{Lieser74,SavrasovJ} and Coulomb parameter $U$ = 4.0~eV
significant improvement was achieved for calculated value of equilibrium
volume comparing with experimental data. The resulting band structure
suggests that Pu has five localized $f$ electrons with
substantial orbital and spin moments values. Another LSDA+U calculation
with GGA correction gave similar results for different Coulomb
parameter values.~\cite{Bouchet00}

In contrast to experimental data, clearly showing nonmagnetic
state of plutonium,~\cite{HeckerPMS,LosAlamos,Lashleycondmat}
antiferromagnetic ground state was found for fcc-Pu in many
calculations.~\cite{Soderlind02,Soderlind01,Wang01,Postnikov00}
The fully relativistic spin-polarized calculation for
fcc-plutonium phase was performed in Ref.~\onlinecite{Solovyev91}.
In another work~(Ref.~\onlinecite{Penicaud97}) it was shown that
`switching off' the hybridization makes better agreement with
experiment for equilibrium volume in $\delta$-Pu. For augmented
plane waves basis (FLAPW),~\cite{Singh94} Kutepov {\it
et~al.}~\cite{Kutepov03} using the fully relativistic
spin-polarized (RSP) method with GGA approximation obtained large
orbital and spin moments. Antiferromagnetic ordered state was
found lower in total energy of the system comparing with
nonmagnetic one for both $\alpha$- and $\delta$-Pu
phases.~\cite{Kutepov03,Kutepov04}

Another scheme was proposed by Eriksson {\it et~al}.~\cite{Eriksson99}
Plutonium 5$f$ electrons were divided into
localized and delocalized ones. Self-interaction correction
SIC-LSDA method\cite{Petit01,Perdew81,Temmerman98,SIC} was used to
calculate volumes and total energies of Pu ion in different
configurations (with different valencies\cite{Petit04,Petit01} of Pu
ion). In contrast to
experiments,~\cite{Moore03} $LS$-coupled ground state was found to
be lower in the total energy for any configuration of Pu
ion\cite{Petit04} and $f^3$ configuration energetically
preferable. For many binary Pu compounds SIC-LSDA calculations
showed that experimental data are better described by the model of
coexisting localized and delocalized electrons.~\cite{Petit02} In
particular, valency +5 was found for Pu ion in PuN and +3 in PuTe
and PuSb.~\cite{Petit02} Pu ion in plutonium dioxide by the same
approach was obtained in $f^4$
configuration.~\cite{Petit04,Petit03} Similar hybrid schemes,
disordered local moment (DLM) method\cite{Niklasson03} gave
qualitative agreement with the photoemission spectrum for
$\delta$-Pu.~\cite{Wills04}

Experiments suggest the presence of significant
correlation effects for 5$f$ electrons in plutonium. Dynamical Mean-Field
Theory\cite{KotliarVollhardt,DMFT,condmat} (DMFT) is a powerful
tool in studying such effects. Merging of LDA-based methods with
DMFT gave a new calculation scheme -- LDA+DMFT
method.~\cite{Anis97,Held03,Savrasov04} Its application to
plutonium problem\cite{Savrasov01,Savrasov04} gave a better
agreement with photoemission experiments and a double minimum
curve of the total energy as a function of volume.

In this paper we present the results of calculation by LDA+U with
spin-orbit coupling (LDA+U+SO) method of electronic
structure and magnetic properties for
metallic plutonium in $\alpha$ and $\delta$ phases
and series of Pu compounds, namely PuN, PuCoGa$_5$, PuRh$_2$,
PuSi$_2$, PuTe, and PuSb. Exchange
interaction (spin polarization) term in the Hamiltonian
was implemented in a general nondiagonal matrix form. This form
is necessary for the correct description of 5$f$ electrons
for the case of $jj$ coupling and intermediate coupling 
schemes which we have found to be
valid in pure metallic Pu and its compounds.
For metallic Pu in both $\delta$ and $\alpha$ phases, our
calculations gave nonmagnetic $f^6$ ground state with six
$f$ electrons in fully occupied $j$ = 5/2 sub-shell and spin,
orbital, and total moments values equal to zero. 
This result is in agreement with experimental data
for metallic plutonium,~\cite{Lashleycondmat} but we have
found magnetic ground state for all Pu-based compounds
investigated in this work. For the present investigation we have 
chosen plutonium compounds with formal Pu ion valency +3 (PuN and 
PuSb) and +2 (PuTe) and also intermetallic compounds where valence
state of Pu ion is not obviously defined. All calculated
Pu-compounds possess magnetic moments on Pu ions with
$f^5$, $f^6$ or mixed configurations. We were interested mainly
in magnetic state problem and discussion of equilibrium volume
and volume transition in plutonium and its compounds could be
found elsewhere.~\cite{Benedict93} A strong competition between
spin-orbit coupling and exchange interactions with a delicate
balance was found to determine magnetic state of Pu and its compounds.

The organization of this paper is as follows. Sec.~\ref{Method}
describes LDA+U with spin-orbit coupling (LDA+U+SO) method used in
the present work. In Sec.~\ref{Pu} results of calculations for
metallic $\delta$- and $\alpha$-Pu phases are discussed in detail.
Then Sec.~\ref{Plutonium compounds} describes  applications of
LDA+U+SO method to various plutonium compounds, including PuN in
Subsec.~\ref{PuN}, PuCoGa$_5$ in Subsec.~\ref{PuCoGa5}, PuRh$_2$
in Subsec.~\ref{PuRh2}, PuSi$_2$ in Subsec.~\ref{PuSi2}, PuTe in
Subsec.~\ref{PuTe}, and PuSb in Subsec.~\ref{PuSb}. Sec.~\ref{Dis}
summarizes the paper.

\section{Method}
\label{Method} A strong spin-orbit coupling in actinides together
with magnetism and Coulomb interactions rise a problem which was
not present for materials without 5$f$ electrons. For 3$d$ and
4$f$ elements compounds relatively weak spin-orbit coupling
results in Russell-Saunders coupling ($LS$ coupling) scheme with
$\mathbf S$ and $\mathbf L$ operators well defined. Then the basis
of $LS$ orbitals, which are eigenfunctions of both spin $\mathbf S$ and
orbital moment $\mathbf L$ operators, is a good choice. In this case it is
possible to define quantization axis in the direction of spin
moment vector so that occupation and potential matrices will be
diagonal in spin variables. When spin-orbit coupling is stronger 
than exchange (Hund)
interaction, $jj$ coupling scheme is valid with a well defined
total moment $\mathbf J$, but not spin $\mathbf S$ and orbital
$\mathbf L$ moments. In this case, the basis of eigenfunctions of
total moment operator $\{jm_j\}$ is a best choice. The matrix of
spin-orbit coupling operator is
diagonal in this basis but not the exchange interaction
(spin-polarization) term in the Hamiltonian.

The situation for 5$f$ electrons is more complicated. The strengths
of spin-orbit coupling and exchange interaction are comparable so
that both aforementioned limits: $LS$ coupling and $jj$ coupling
schemes are not valid and intermediate coupling scheme is
needed to describe 5$f$ shell in actinides. In this case
occupation matrix is diagonal neither in $\{LS\}$, nor in
$\{jm_j\}$ orbital basis and both terms in the Hamiltonian:
spin-orbit coupling and exchange interaction, must be taken in a
general nondiagonal matrix form.

In LDA+U method\cite{LDA+U} the energy functional depends, in
addition to the charge density $\rho (\mathbf{r)}$, on the
occupation matrix $n_{mm^{\prime }}^{s s^{\prime }}$ (LDA+U method
in general nondiagonal in spin variables form was defined
in Ref.~\onlinecite{Solovyev98}):
\begin{equation}
E^{LDA+U}[\rho (\mathbf{r}),\{n\}]=E^{LDA}[\rho (\mathbf{r)}%
]+E^U[\{n\}]-E_{dc}[\{n\}]  \label{U1}
\end{equation}
where $\rho (\mathbf{r})$ is the charge density and $E^{LDA}[\rho
(\mathbf{r})]$ is the standard LDA functional. The occupation
matrix is defined as:
\begin{equation}
n_{mm^{\prime }}^{s s^{\prime }}=-\frac 1\pi
\int^{E_F}ImG_{mm^{\prime }}^{s s^{\prime }}(E)dE \label{Occ}
\end{equation}
where $G_{mm^{^{\prime }}}^{s s^{\prime }}(E)=\langle inlms \mid
(E-\widehat{H}_{LDA+U})^{-1}\mid inlm^{\prime }s^{\prime
}\rangle $ are the elements of the Green function matrix in local
orbital basis set ($i$ -- atomic index, $n$ -- principal quantum
number, $lm$ -- orbital quantum numbers, and $s$ -- spin index). In
the present work this basis set was formed by LMT-orbitals from
LMTO method.~\cite{LMTO}
In Eq. (\ref{U1}) Coulomb interaction energy $E^U[\{n\}]$ term is a
function of occupancy matrix $n_{mm^{\prime }}^{s s^{\prime }}$:
\begin{equation}
\begin{array}{c}
E^U[\{n\}]=\frac 12\sum_{\{m\},ss^{\prime }}\{\langle
m,m^{\prime \prime }\mid V_{ee}\mid m^{\prime },m^{\prime \prime \prime
}\rangle n_{mm^{\prime }}^{ss}n_{m^{\prime \prime }m^{\prime
\prime \prime }}^{s^{\prime }s^{\prime }} \\
\\
-\langle m,m^{\prime \prime }\mid V_{ee}\mid m^{\prime \prime
\prime },m^{\prime }\rangle )n_{mm^{\prime }}^{ss^{\prime
}}n_{m^{\prime \prime }m^{\prime \prime \prime }}^{s^{\prime }s
}\}
\end{array}
\label{upart}
\end{equation}
where $V_{ee}$ is the screened Coulomb interaction between the $nl$
electrons. Finally, the last term in Eq. (\ref{U1}) correcting for
double counting is a function of the total number of electrons in
the spirit of LDA which is a functional of total charge density:
\begin{equation}
E_{dc}[\{n\}]=\frac {1}{2}UN(N-1)-\frac {1}{4}JN(N-2) 
\label{U3}
\end{equation}
where $N$ = $Tr(n_{mm^{\prime }}^{ss^{\prime }})$ is a total number of
electrons in the $nl$ shell. $U$ and $J$ are screened Coulomb and
Hund exchange parameters which could be determined in constrain
LDA calculations.~\cite{Gunnarsson89,Anisimov91a} The latter ($J$)
in this paper is everywhere denoted as $J_H$ not to be confused
with total moment operator $J$. The screened Coulomb interaction
matrix elements $\langle m,m^{\prime \prime }\mid V_{ee}\mid
m^{\prime },m^{\prime \prime \prime }\rangle$ could be expressed
via parameters $U$ and $J_H$ (see Ref.~\onlinecite{LDA+U}).

The functional Eq. (\ref{U1}) defines the effective single-particle
Hamiltonian with an orbital dependent potential added to the
usual LDA potential:
\begin{equation}
\widehat{H}_{LDA+U}=\widehat{H}_{LDA}+\sum_{ms,m^{\prime
}s^{\prime }}\mid inlms \rangle V_{mm^{\prime }}^{ss^{\prime
}}\langle inlm^{\prime }s^{\prime }|  \label{hamilt}
\end{equation}
\begin{equation}
\begin{array}{c}
V_{mm^{\prime }}^{ss^{\prime }}=\delta _{ss^{\prime
}}\sum_{m^{\prime \prime},m^{\prime \prime \prime}}\{\langle m,m^{\prime \prime }\mid V_{ee}\mid m^{\prime
},m^{\prime \prime \prime }\rangle n_{m^{\prime \prime }m^{\prime \prime
\prime }}^{-s,-s }+ \\
\\
(\langle m,m^{\prime \prime }\mid V_{ee}\mid m^{\prime },m^{\prime \prime \prime }\rangle
-\langle m,m^{\prime \prime }\mid V_{ee}\mid m^{\prime \prime \prime },
m^{\prime }\rangle )n_{m^{\prime \prime }m^{\prime \prime \prime }}^{ss}\}- \\
\\
\left( 1-\delta _{ss^{\prime }}\right) \sum_{m^{\prime \prime},m^{\prime \prime \prime}}\langle
m,m^{\prime \prime }\mid V_{ee}\mid m^{\prime \prime \prime },m^{\prime
}\rangle )n_{m^{\prime \prime }m^{\prime \prime \prime }}^{s^{\prime
}s} \\
\\
-U(N-\frac{1}{2})+\frac{1}{2}J(N-1).
\end{array}
\label{Pot}
\end{equation}

In this paper we used method abbreviated as LDA+U+SO which
comprises non-diagonal in spin variables LDA+U Hamiltonian 
Eq. (\ref{Pot}) with spin-orbit (SO) coupling 
realized on the base of TB-LMTO-ASA code:~\cite{LMTO}
\begin{equation}
\begin{array}{c}
\widehat{H}_{LDA+U+SO}=\widehat{H}_{LDA+U}+\widehat{H}_{SO},\\
\\ 
\widehat{H}_{SO}=\lambda \cdot  \widehat{\mathbf L} \cdot
\widehat{\mathbf S} 
\label{eq:LS}.
\end{array}
\end{equation}
In $LS$ basis spin-orbit (SO) coupling matrix has
diagonal $(H_{SO})^{s,s}_{m^{\prime},m}$ as well as off-diagonal 
$(H_{SO})^{\uparrow,\downarrow}_{m^{\prime},m}$ and 
$(H_{SO})^{\downarrow,\uparrow}_{m^{\prime},m}$
non-zero matrix elements (in complex spherical harmonics): 
\begin{equation}
\begin{array}{c}
(H_{SO})^{\uparrow,\downarrow}_{m^{\prime},m} =
\frac{\lambda}{2}
\sqrt{(l+m)(l-m+1)}(\delta_{m^\prime,m-1})\\
\\
(H_{SO})^{\downarrow,\uparrow}_{m^{\prime},m} =
\frac{\lambda}{2}\sqrt{(l+m)(l-m+1)}(\delta_{m^\prime-1,m}) \\
\\
(H_{SO})^{s,s}_{m^{\prime},m} = \lambda m s \delta_{m^\prime,m}
\end{array}
\label{SO-nondiag}
\end{equation}
where $s$ = +1/2, --1/2.~\cite{LandauSO}

The exchange interaction is originally a part of the Coulomb
interaction (a second term in Eq. (\ref{upart})). In the case when 
spin moment operator $\mathbf S$ is well defined, so that
density and potential matrices can be made diagonal in spin variables, it
is convenient to define spin-polarized density and spin polarized
potential. In a most general form exchange interaction allowing
non-collinearity of the spin polarization in space was introduced
in density functional theory\cite{DFT} (DFT) by von Barth {\it et
al.}~\cite{vonBarth72,Sandratskii98}
and used later in Ref.~\onlinecite{Kubler88,Nordstrom96}:
\begin{equation}
\lbrace [-\nabla ^2+V_{ext}(r)+\int d^3r^{\prime  }n(r^{\prime })%
\frac{e^2}{\mid r-r^{\prime }\mid }+v_{xc}(r;n,\vec{m} )]\sigma _o
+\vec b_{xc}(r;n,\vec{m}) \cdot
 \vec\sigma \rbrace \Psi_i(r)
=\varepsilon_i\Psi_i(r)
\end{equation}
where $\Psi_i(r)$ are one-electron spinors, $\sigma_i$ are Pauli
matrices. Exchange-correlation potentials determined as
variational derivatives over space charge density $n(r)$ or
magnetization vector $\vec m(r)$ are:
\begin{equation}
 v_{xc}(r)=\frac {\delta E_{xc}[n,\vec{m}]} {\delta n(r)},~~~
 \vec b_{xc}(r)=\frac {\delta E_{xc}[n,\vec{m}] } {\delta \vec m(r)}=
 \frac {\delta E_{xc}[n,m]} {\delta m} \hat m(r).
\end{equation}
The exchange interaction is described here by the local magnetic
field $\vec b_{xc}$ which acts on all orbitals in the same way.
The potential matrix in $LS$ basis for this field is diagonal
in spin variables if quantization axis is chosen along direction
of $\vec b_{xc}$.

A strong spin-orbit coupling in 5$f$ shell produces an occupation
matrix Eq. (\ref{Occ}) and the corresponding potential matrix
Eq. (\ref{Pot}) which can not be made diagonal in spin variables by
rotation of the quantization axis. Such rotation must be done to
the direction of the spin moment vector, so that only $S_z$
components were present while $S_x,S_y$ components (responsible
for off-diagonal in spin variable terms in Hamiltonian) set to
zero. This is possible in the case where $LS$ coupling scheme is
valid and spin moment operator $\mathbf S$ is well defined, which
is true for 3$d$ and 4$f$ elements compounds, but not for 5$f$
electrons of actinides. For later materials not $LS$ coupling
scheme is valid but $jj$ coupling or intermediate coupling 
schemes where spin moment operator $\mathbf S$ is not well
defined.

That means that the widely used\cite{Solovyev91,Kutepov03} form of
spin-polarized potential based on Local Spin Density Approximation
(LSDA) can not be applied for actinides, because LSDA neglects
off-diagonal in spin variables terms in both spin density and
potential. Those terms appear due to spin-orbit coupling
Eq. (\ref{SO-nondiag}) and suppress spin moment formation while
exchange (Hund) part of Coulomb interaction prefers fully
spin-polarized state with diagonal in spin variables occupation
and potential matrices. A competition between spin-orbit coupling
and exchange interaction determines the magnetic state of 5$f$
shell. Neglecting off-diagonal in spin variables potential terms
in the methods using LSDA potential enhances the tendency to spin
moment formation. (In Sec.~\ref{Pu} we show that neglecting
off-diagonal in spin variables potential terms in LDA+U+SO method
results in a very strong increase of calculated spin and orbital
moment values.) This fact could explain why antiferromagnetic 
ground state with a large value of spin moment was obtained 
in fully relativistic spin-polarized method 
calculations\cite{Solovyev91,Kutepov03} in disagreement 
with experimental data.~\cite{Lashleycondmat,HeckerPMS,LosAlamos}

\section{Metallic plutonium}
\label{Pu}
\subsection{$\delta$ phase of plutonium}
\label{Delta-Pu}
\begin{figure}[!t]
\begin{center}
\epsfxsize=14cm
\epsfbox{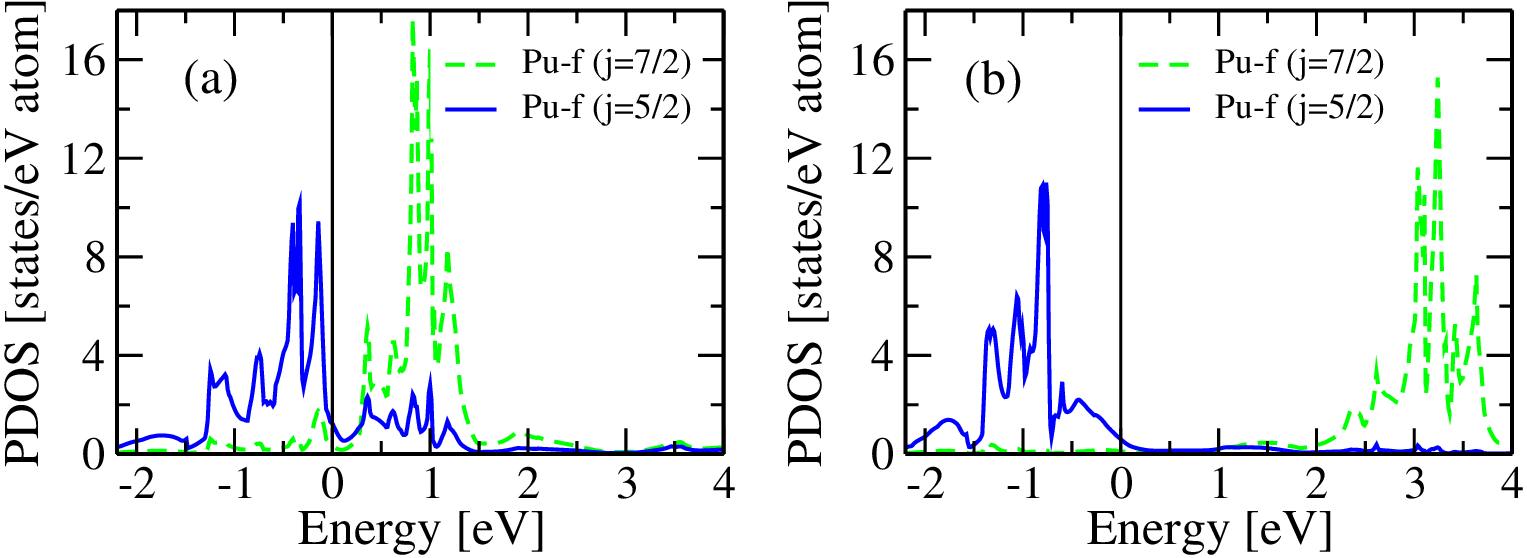}
\end{center}
\caption{ Partial $f^{5/2}$ and $f^{7/2}$ contributions to 5$f$ band
densities of states of $\delta$-plutonium calculated with LDA (a)
and LDA+U+SO method (b).}
\label{fig:deltaPuDOS}
\end{figure}
Phase diagram of plutonium is the most complex of all
elements.~\cite{Young01} Of especial interest is fcc $\delta$ phase of Pu
($a$ = 4.636~\AA). A phase transition from $\delta$- phase to
low-temperature monoclinic $\alpha$-plutonium phase ($a$ = 6.183~\AA,
$b$ = 4.824~\AA, $c$ = 10.973~\AA,~and $\beta$ = 101.79$^\circ$) is
accompanied with a very large (19\%) volume contraction. In the present
work we were interested in the problem of magnetic state for Pu metal
in $\alpha$ and $\delta$ phases.

In the LDA+U calculation scheme the values of direct ($U$) and
exchange ($J_H$) Coulomb parameters must be determined as a first
step of calculation procedure. It can be done in {\it{ab-initio}}
way via constrain LDA
calculations.~\cite{Anisimov91a,Gunnarsson89} In our calculations
Hund exchange parameter $J_H$ was found to be $J_H$ = 0.48~eV, 
much smaller than the value used in previous LDA+U
studies.~\cite{SavrasovJ} Coulomb repulsion parameter value $U$ = 3.84~eV 
was obtained in a good agreement with previously used $U$ = 
4~eV.~\cite{Dedericks84,Desclaux84,Turchi99,Savrasov00,Savrasov01,Dai03}
Coulomb interaction parameters describe a {\it{screened}} Coulomb
interaction between 5$f$ electrons and hence crucially depend on
the channels of screening taken into account in constrain LDA
calculations. That is true for the direct Coulomb parameter $U$,
but exchange Coulomb parameter $J_H$ corresponds to the
{\it{difference}} of interaction energy for the electrons pairs
with the opposite and the same spin directions. As the screening
process is defined by the charge but not spin state of the ion,
the screening contribution is cancelled for exchange Coulomb
interaction and parameter $J_H$ does not depend on the screening
channels choice. For example for 3$d$ elements compounds, direct
Coulomb interaction parameter varies from 8~eV for late transition
metal oxides (NiO and CuO) to 3~eV for the beginning of the row
(Ti and V oxides). At the same time exchange Coulomb parameter
$J_H$ is in the range of 0.85-0.95~eV for the whole 3$d$ row. In
our constrain LDA calculations we have found that $J_H$ value is
equal 0.48~eV within the accuracy 0.01~eV for both metallic Pu
phases and for all plutonium compounds investigated in the present
work.

In our constrain LDA calculations it was supposed that $f$ shell
is screened only by $s$, $p$, and $d$ electrons, so that 5$f$
electrons were considered to be completely localized and
not participating in the screening. LDA calculations with
spin-orbit coupling taken into account show
(Fig.~\ref{fig:deltaPuDOS}(a)) well separated subbands corresponding
to the total moment values $j$ = 5/2 and $j$ = 7/2 with the Fermi level
crossing the top of $j$ = 5/2 band. This
splitting of 5$f$ shell into nearly filled $f^{5/2}$ and empty
$f^{7/2}$ subshells can justify an additional screening of
$f^{5/2}$ electrons by $f^{7/2}$ orbitals. In the constrain LDA
calculations with this screening channel taken into account, 
Coulomb parameter value was obtained as $U$ = 0.44~eV.

\begin{figure}[!t]
\begin{center}
\epsfxsize=7cm
\epsfbox{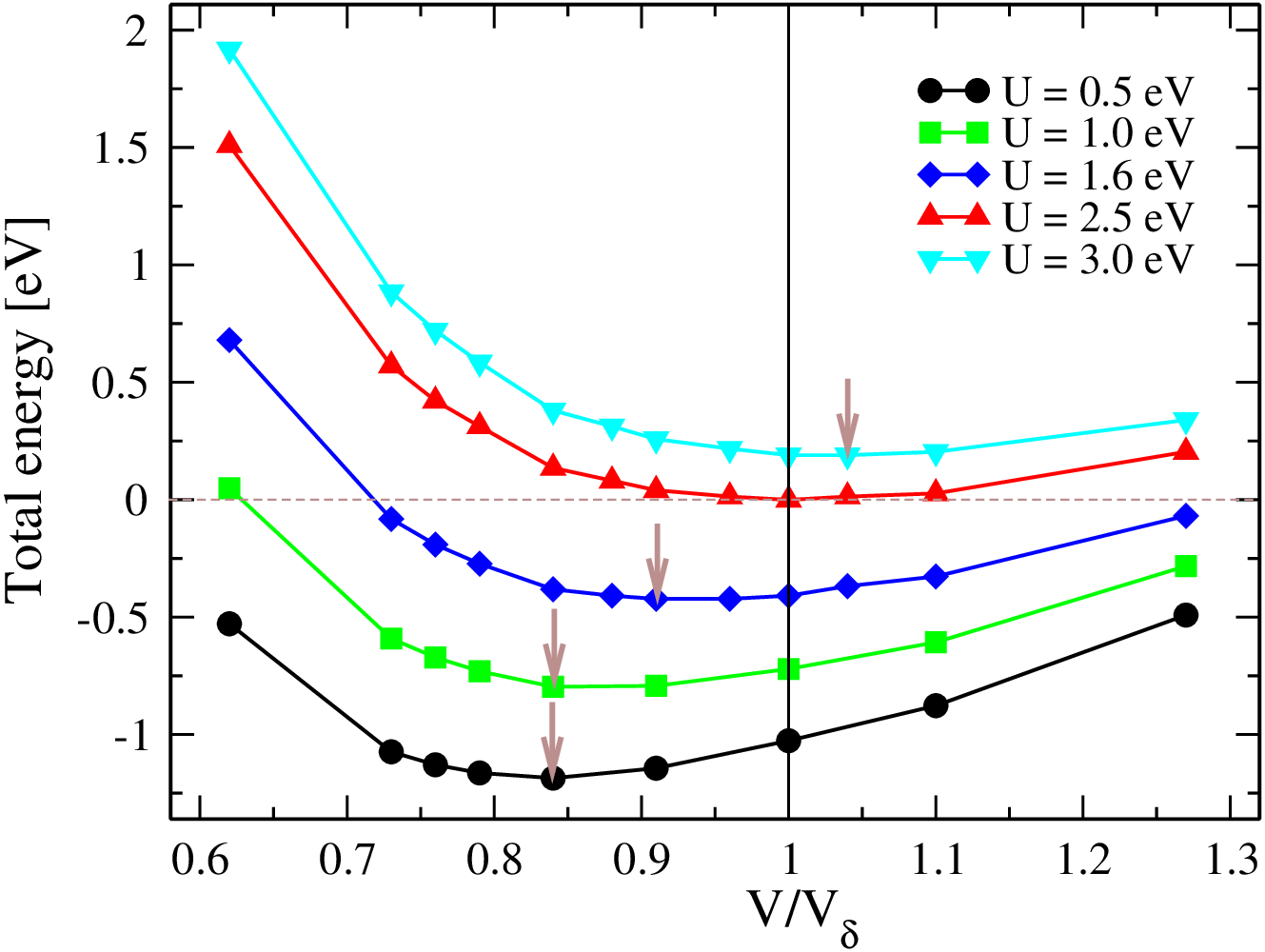}
\end{center}
\caption{Total energy as a function of volume for fcc-Pu calculated with various
Coulomb parameter $U$ values (V$_{\delta}$ corresponds to experimental volume value
for Pu in $\delta$ phase). Arrows indicate minima of the curves giving equilibrium
volume values. Calculated equilibrium volume is equal to experimental value for 
the curve with $U$ = 2.5~eV.} 
\label{fig:volU}
\end{figure}

Plutonium is considered to be on the border between
completely localized 5$f$ electrons of americium and itinerant
nature of them for early actinides, so that both limits,
completely localized and totally itinerant, are not appropriate
for Pu 5$f$ electrons. The correct value of $U$ should be somewhere 
in between the values calculated in those two limits (3.84~eV without 
any $f$ electrons participating in the screening and 0.44~eV for full
$f^{7/2}$-screening). As an additional requirement to determine $U$
value we have chosen an equality of the calculated equilibrium
volume of fcc-Pu to the experimental value of $\delta$ phase (see
Fig.~\ref{fig:volU}). This requirement is fulfilled for $U$ = 2.5~eV
and we used this Coulomb interaction parameter value in LDA+U+SO
calculations for metallic Pu in both $\alpha$ and $\delta$ phases
and for all Pu compounds investigated in this work.
\begin{table}[!t]
\caption{Electronic configuration of Pu ion 5$f$ shell in
$\delta$-plutonium calculated with LDA+U+SO method ($U$ = 2.5~eV)
for various values of Hund exchange parameter $J_H$. The largest
values of occupation matrices off-diagonal elements (OD) in
$\{LS\}$ and $\{jm_j\}$ basis sets are given in the second and
third columns. The seven largest eigenvalues of occupation matrix
are presented from fourth to tenth columns. Columns from eleven to
thirteen show the calculated values for spin ($S$), orbital ($L$),
and total ($J$) moments.~\cite{signnote} Last four columns contain
partial contributions of $f^5$ and $f^6$ configurations and types
of coupling for 5$f$ shell of plutonium ion (see the text for
explanations).}
\begin{center}
\begin{tabular}{lllllllllll}
\hline \hline
$J_H$,~eV & OD$_{\{LS\}}$ & OD$_{\{jm_j\}}$ & largest eigenvalues & %
$S$ & $L$ & $J$ & $f^5$ & $f^6$ & $jj$ & $LS$\\
\hline
0.40 & 0.431 & 0.022 & 0.041 0.885 0.885 0.910 0.910 0.910 0.910 &   0   &   0   &   0   &   0  & 1.00 & 1.00 & 0.00\\
0.43 & 0.432 & 0.021 & 0.041 0.886 0.886 0.911 0.911 0.911 0.911 &   0   &   0   &   0   &   0  & 1.00 & 1.00 & 0.00\\
0.48 & 0.433 & 0.008 & 0.040 0.889 0.889 0.912 0.912 0.912 0.912 &   0   &   0   &   0   &   0  & 1.00 & 1.00 & 0.00\\
0.49 & 0.435 & 0.019 & 0.040 0.889 0.890 0.912 0.912 0.913 0.914 & 0.083 & 0.099 & 0.016 & 0.01 & 0.99 & 0.97 & 0.03\\
0.50 & 0.433 & 0.127 & 0.041 0.880 0.884 0.909 0.918 0.918 0.925 & 0.582 & 0.692 & 0.110 & 0.04 & 0.96 & 0.81 & 0.19\\
0.55 & 0.412 & 0.279 & 0.045 0.838 0.884 0.912 0.929 0.938 0.945 & 1.369 & 1.640 & 0.271 & 0.11 & 0.89 & 0.54 & 0.46\\
0.60 & 0.392 & 0.346 & 0.050 0.823 0.888 0.916 0.935 0.949 0.956 & 1.746 & 2.064 & 0.318 & 0.13 & 0.87 & 0.41 & 0.59\\
\hline \hline
\end{tabular}
\end{center}
\label{tab:deltaPua}
\end{table}
\begin{table}[!t]
\caption{Electronic configuration of Pu ion in $\delta$-plutonium calculated
with LDA+U+SO method as a function of the 5$f$ band shift value $\Delta$.
(Notations are the same as for Table~\ref{tab:deltaPua}.)}
\begin{center}
\begin{tabular}{lllllllllll}
\hline \hline
$\Delta$,~eV & OD$_{\{LS\}}$ & OD$_{\{jm_j\}}$ & largest eigenvalues & %
$S$ & $L$ & $J$ & $f^5$ & $f^6$ & $jj$ & $LS$\\
\hline
  0  & 0.433 & 0.008 & 0.040 0.889 0.889 0.912 0.912 0.912 0.912 &   0   &   0   &   0   &   0  & 1.00 & 1.00 & 0.00\\
0.14 & 0.431 & 0.012 & 0.040 0.879 0.879 0.906 0.906 0.906 0.906 & 0.023 & 0.029 & 0.006 &   0  & 1.00 & 0.99 & 0.01\\
0.27 & 0.431 & 0.044 & 0.040 0.864 0.866 0.897 0.900 0.903 0.905 & 0.210 & 0.286 & 0.076 & 0.03 & 0.97 & 0.93 & 0.07\\
0.41 & 0.427 & 0.188 & 0.042 0.642 0.865 0.919 0.919 0.934 0.942 & 0.920 & 1.700 & 0.780 & 0.31 & 0.79 & 0.70 & 0.30\\
0.82 & 0.432 & 0.256 & 0.043 0.254 0.910 0.924 0.942 0.954 0.959 & 1.316 & 3.040 & 1.724 & 0.69 & 0.31 & 0.56 & 0.44\\
1.36 & 0.419 & 0.256 & 0.040 0.086 0.886 0.886 0.938 0.947 0.962 & 1.378 & 3.609 & 2.231 & 0.89 & 0.11 & 0.53 & 0.47\\
\hline \hline
\end{tabular}
\end{center}
\label{tab:deltaPub}
\end{table}

The LDA+U+SO calculations for metallic Pu in $\delta$ phase gave a
nonmagnetic ground state with zero values of spin $\mathbf S$,
orbital $\mathbf L$, and total $\mathbf J$ moments. The occupation
matrix Eq. (\ref{Occ}) has six eigenvalues close to unit (see in
Table~\ref{tab:deltaPua} the row corresponding to $J_H$ = 0.48~eV)
and is nearly diagonal in the $\{jm_j\}$ basis of eigenfunctions
of total moment $\mathbf J$. That gives a $f^6$ configuration of
Pu ion in $jj$ coupling scheme. In the Fig.~\ref{fig:deltaPuDOS}(b)
the partial densities of states for $f^{5/2}$ and $f^{7/2}$
subshell orbitals are presented. In agreement with the occupation
matrix analysis one can see nearly completely filled $f^{5/2}$
band with the Fermi level on the top of it and an empty $f^{7/2}$
band. The separation between the centers of these bands is
$\approx$ 4~eV.

The origin of this nonmagnetic LDA+U+SO solution can be traced to
the results of standard LDA calculations without Coulomb
correlation correction but with spin-orbit coupling taken into account 
(see Fig.~\ref{fig:deltaPuDOS}(a)). The 5$f$ band is
split by spin-orbit coupling into well pronounced $f^{5/2}$ and
$f^{7/2}$ bands with a separation between their centers $\approx$ 
1.5~eV. Fermi level is in the pseudogap between subbands closer to
the top of $f^{5/2}$ band. Comparing Fig.~\ref{fig:deltaPuDOS}(a)
and Fig.~\ref{fig:deltaPuDOS}(b) one can see that taking into
account Coulomb correlations via LDA+U correction potential
(Eq. \ref{Pot}) does not change qualitatively the band structure,
only the separation between subbands increases from 1.5~eV
to 4~eV according to the value of $U$ = 2.5~eV. Another effect
of Coulomb interaction is nearly pure orbital character
of $f^{5/2}$ and $f^{7/2}$ bands in LDA+U+SO calculations
comparing with a significant admixture of $f^{5/2}$ orbitals 
to the nominally $f^{7/2}$ band and vice versa in the LDA results 
(Fig.~\ref{fig:deltaPuDOS}(a)). One can say, that
nonmagnetic $J_H$ = 0 solution with $f^{5/2}$ subshell filled with six
electrons and empty $f^{7/2}$ band is already `preformed' in LDA. Role 
of Coulomb correlations in LDA+U method is to make it more pronounced 
with a pure orbital nature of the bands and increased energy 
separation between them.
\begin{figure}[!t]
\begin{center}
\epsfxsize=7cm
\epsfbox{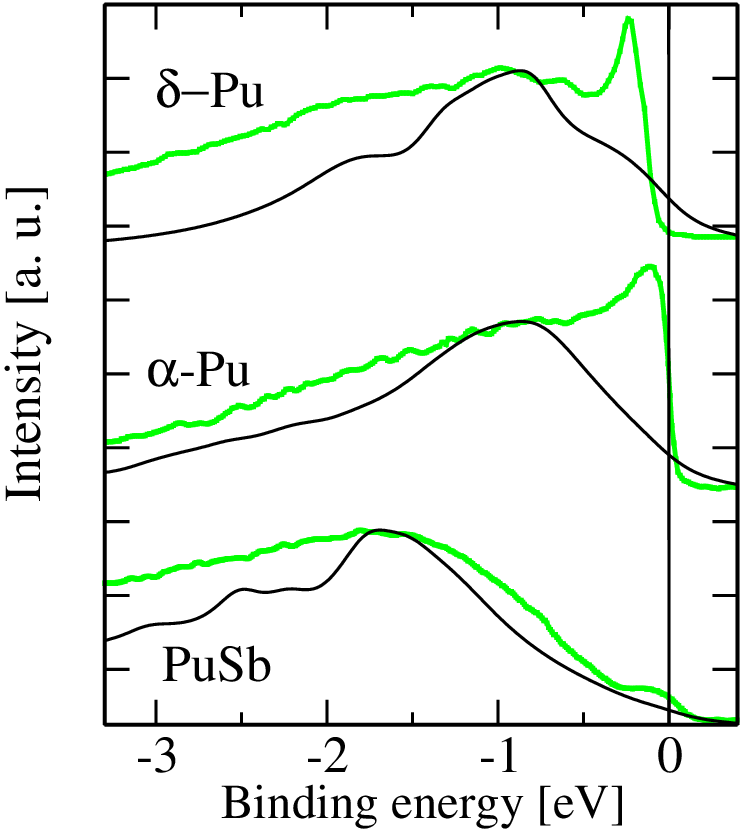}
\end{center}
\caption{Experimental (PES spectra for $\delta$- and $\alpha$-plutonium
with $h\nu$ = 41~eV and T = 80~K\cite{PESa} and angle-integrated PES 
for PuSb at 15~K\cite{Durakiewicz04}) and calculated (thin line) 
photoemission spectra for metallic Pu in $\delta$ and $\alpha$ phases 
and for PuSb.}
\label{UPS}
\end{figure}

Photoemission spectroscopy is a direct tool for investigating
electronic structure. Total density of states obtained in our
LDA+U+SO study were multiplied with the Fermi function
corresponding to the temperature of experiment and broadened with
0.2~eV Lorentian to account for the instrumental resolution (see
Fig.~\ref{UPS}). With the exception of the sharp peak near the
Fermi level an overall agreement between experimental and
calculated spectra for $\delta$-Pu is satisfactory. This peak is
usually considered as a sign of strong many-body effects, which
can be described by Dynamical Mean-Field
Theory\cite{KotliarVollhardt,DMFT,condmat} (DMFT) (see
Ref.~\onlinecite{Savrasov01,Savrasov04}) but not by static
mean-field approximation which is a basis of LDA+U method.
However, LDA+U can reproduce the lower and upper Hubbard bands, on
which partially filled $f$ band is split by Coulomb interaction.
Indeed the peak in experimental spectrum at $\approx$ 1~eV
corresponding to the lower Hubbard band is well described by the
calculated spectrum.

Our results are in agreement with experimental data showing the
absence of ordered or disordered magnetic moments for plutonium in
both $\delta$ and $\alpha$ phases.~\cite{Lashleycondmat,HeckerPMS,LosAlamos}
However, all previous electronic structure calculations gave a magnetic
(usually antiferromagnetic) state lower in energy than a nonmagnetic state.
To clarify this problem we have carried out an investigation of
the stability of our nonmagnetic ground state to the parameters of
the calculations. The influence on it of the different
approximations for exchange interaction and spin-orbit coupling
terms in the Hamiltonian was investigated as well.

As a first step we have varied the value of exchange Coulomb
parameter $J_H$ around $J_H$ = 0.48~eV value obtained in our
constrained LDA calculations. In Table~\ref{tab:deltaPua} the
results of LDA+U+SO calculations with $J_H$ values in the range
from 0.40~eV till 0.60~eV are presented. One can see that even
15\% increasing of $J_H$ value is enough to result in a magnetic
ground state with a sizable values of spin and orbital moments.
However, the total moment value remains very small, less than
0.3~$\mu_B$.

In addition to moment values, we show in the
Table~\ref{tab:deltaPua} values of the largest off-diagonal
elements of occupation matrix Eq. (\ref{Occ}) in $\{LS\}$ and
$\{jm_j\}$ basis. As one can see, for the value
$J_H$ = 0.48~eV obtained in constrain LDA calculations,
$\{jm_j\}$ basis is the most appropriate with largest off-diagonal 
elements values less than 0.01, while in $\{LS\}$ basis these
values are very large (more than 0.4). That can be interpreted as a
pure case of $jj$ coupling with a well defined total moment ($\mathbf J$)
but not spin ($\mathbf S$) and orbital ($\mathbf L$) moments.
With increasing of $J_H$ value the off-diagonal elements in $\{jm_j\}$
basis grow fast and for $J_H$ = 0.60~eV become comparable with 
the off-diagonal elements in $\{LS\}$ basis making $jj$ coupling 
scheme not valid any more. As the off-diagonal elements 
in $\{LS\}$ basis are still very large ($\approx$ 0.4), 
the $LS$ coupling scheme is also not appropriate 
and intermediate coupling scheme is needed to describe 
the magnetic solution.

The eigenvalues of occupation matrix Eq. (\ref{Occ}) presented in
Table~\ref{tab:deltaPua} show six orbitals with occupancies close
to unit for all values of $J_H$. Considering this one can conclude 
that Pu ion is predominantly in $f^6$ configuration in magnetic 
as well as in nonmagnetic state. Only nonmagnetic ground states 
corresponding to the values of Hund parameter $J_H\leq$ 0.48~eV can
be interpreted as a pure $f^6$ configuration in $jj$ coupling
scheme. Magnetic solutions for $J_H>$ 0.48~eV correspond to
predominantly $f^6$ configuration in intermediate coupling scheme.
\begin{figure}[!t]
\begin{center}
\epsfxsize=14cm
\epsfbox{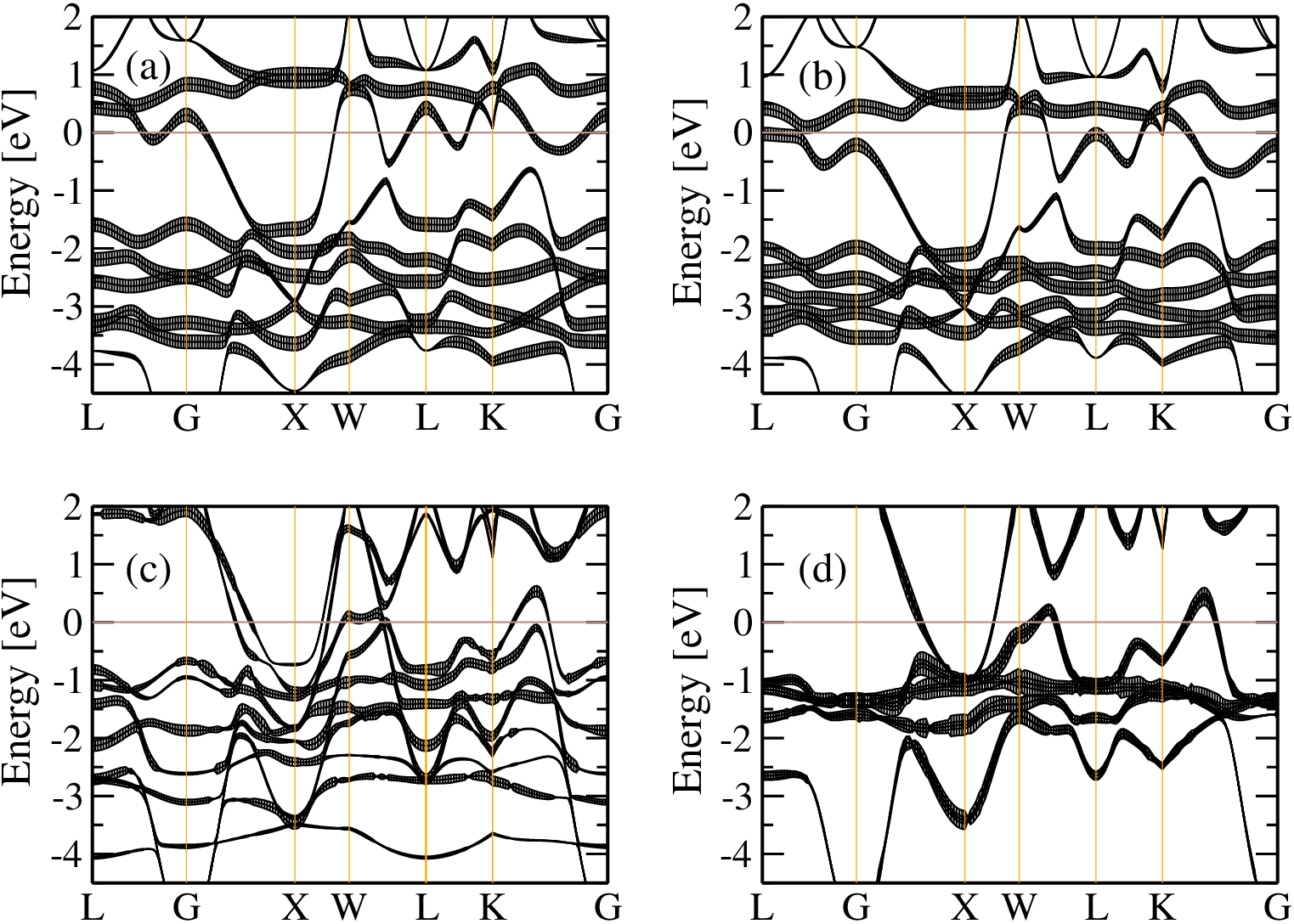}
\end{center}
\caption{(a) Band structure of $\delta$-Pu (the width of the line
shows the contribution of the 5$f$ states to the particular band)
from LSDA+U calculations with Coulomb parameters values
$U$ = 4~eV and $J_H$ = 0.85~eV. (b) The same as (a) but for $J_H$ = 0.48~eV.
(c) The same as (a) but by LDA+U+SO method ($U$ = 4~eV and
$J_H$ = 0.85~eV). (d) Same as (c) but for $J_H$ = 0.48~eV (see text).}
\label{fig:deltaPuBANDSa}
\end{figure}
\begin{table}
\caption{Calculated values for spin, orbital and total moments
from LDA+U+SO calculations with off-diagonal in spin variables
elements in potential matrix (Eq. \ref{Pot}) set to zero. For
comparison the results of LSDA and LSDA+U calculations are
presented.~\cite{signnote}}
\begin{center}
\begin{tabular}{llll}
\hline \hline
Method & $S$ & $L$ & $J$\\
\hline LDA+U+SO ($U$ = 2.5~eV, $J_H$ = 0.48~eV)
& 2.43 & 3.57 & 1.14\\
LSDA+U (Ref.~\onlinecite{Bouchet00})
& 2.50 & 3.70 & 1.20\\
LSDA+U (Ref.~\onlinecite{Savrasov01})
& 2.55 & 3.90 & 1.45\\
\hline LDA+U+SO ($U$ = $J_H$ = 0.48~eV)
& 1.70 & 2.30 & 0.60\\
LSDA (Ref.~\onlinecite{Kutepov03})
& 2.04 & 1.75 & 0.29\\
LSDA (Ref.~\onlinecite{Bouchet00})
& 2.11 & 1.94 & 0.17\\
LSDA (Ref.~\onlinecite{Solovyev91})
& 2.25 & 2.40 & 0.15\\
\hline \hline
\end{tabular}
\end{center}
\label{tab:compcalc}
\end{table}

In order to estimate, how close is a calculated state to one of
the pure states ($f^6$ or $f^5$ configurations, in $jj$ or $LS$
coupling schemes) we propose the following simple formula using
calculated values of spin ($S$), orbital ($L$), and total ($J$) moments. 
Total moment value is the same in both coupling schemes
($jj$ or $LS$): $J$ = 0 for $f^6$ and $J$ = 5/2 for $f^5$. If we
have a mixed state $(1-x)\cdot f^6+x\cdot f^5$ then $x$ can be
defined as $x$ = $J$/2.5. Spin and orbital moment values for $f^6$
configuration are equal to: $S$ = 0, $L$ = 0 in $jj$ coupling
scheme and $S$ = 3, $L$ = 3 in $LS$ coupling scheme. For $f^5$
configuration $S$ = 5/14 $\approx$ 0.36, $L$ = 20/7 $\approx$ 2.86
in $jj$ coupling scheme and $S$ = 5/2, $L$ = 5 in $LS$ coupling
scheme. One can define a mixed coupling scheme with a contribution
of $jj$ coupling equal to $y$ and of $LS$ coupling correspondingly
to $(1-y)$. In this scheme the calculated values of orbital and
spin moment will be equal to
\begin{eqnarray}
L=x\cdot (2.86\cdot y +5\cdot (1-y))+(1-x)\cdot (0\cdot y
+3\cdot (1-y))\\
 S=x\cdot (0.36\cdot y +2.5\cdot
(1-y))+(1-x)\cdot (0\cdot y +3\cdot (1-y)
 \label{LS-model}
\end{eqnarray}
These formulas allow to determine the coefficient $y$.

The values of coefficients $x$ and $y$ calculated in this way are
shown in Table~\ref{tab:deltaPua}. For values of Hund parameter
$J_H\leq$ 0.48~eV they correspond to the pure $f^6$ configuration
in 100\% $jj$ coupling scheme. With increase of $J_H$ value the
contribution of $LS$ coupling increases and for $J_H$ close to
0.60~eV both coupling scheme gave approximately equal contribution
demonstrating a clear case of intermediate coupling scheme.
However the configuration is still predominantly $f^6$ with not
more than 10\% admixture of $f^5$.

Not only the strength of exchange interaction parameter $J_H$ is
crucial for the magnetism of the 5$f$ shell in metallic Pu. The
Fermi level is on the top of $f^{5/2}$ band (see
Fig.~\ref{fig:deltaPuDOS}(a)) and a small shift of relative energy position
of 5$f$ and other states can lead to redistribution of
electrons between 5$f$ and $spd$-bands and hence change the
$f$ configuration and magnetic state of Pu ion. To investigate this effect
we run the LDA+U+SO calculations with a constant positive
potential $\Delta$ acting on 5$f$ electrons, which should result in a
charge flow from 5$f$ bands to $spd$-bands (see
Table~\ref{tab:deltaPub}). Already relatively small value of the
shift ($\Delta$ = 0.41~eV) is enough to produce a well pronounced magnetic
ground state with a sizable admixture of $f^5$ configuration
(around one third). For the shift value equal to 1.36~eV the Pu ion
is in a pure $f^5$ configuration with a very large values of spin,
orbital, and total moments.

These constrain LDA+U+SO calculations demonstrate that a relative
position of 5$f$ and other bands is of great importance for the
resulting configuration and magnetic state of Pu ion. Later we
will show that in all investigated in this work plutonium
compounds we obtained a magnetic ground state in contrast to the
nonmagnetic one for metallic Pu.

In order to study the influence of the different approximations
for exchange interaction term in the Hamiltonian we have performed 
at first LSDA+U (electron density and LDA potential were used in 
spin-polarized forms) calculations with the
same Coulomb interaction parameters values as in Ref.~\onlinecite{Savrasov00} 
(exchange parameter value corresponds
to $J_H$ = 0.85~eV\cite{SavrasovJ} and $U$ = 4~eV). In this calculation
the strong magnetic state with a configuration close to $f^5$ was obtained
in agreement with results of Ref.~\onlinecite{Savrasov00}. The 5$f$ bands
dispersion is presented in Fig.~\ref{fig:deltaPuBANDSa}(a) (the width
of the band curve is proportional to the contribution of 5$f$ states
to the band). It agrees well with the corresponding figure in
Ref.~\onlinecite{Savrasov00}. Decreasing of the Coulomb exchange
parameter value $J_H$ from 0.85~eV to the value 0.48~eV (obtained
in our constrain LDA calculations) gave still magnetic solution
with spin $S$ = 2.5 but the 5$f$ configuration is now more close
to $f^6$ than to $f^5$ (the Fermi level (Fig.~\ref{fig:deltaPuBANDSa}(b))
is now above the top of the band which was cut by the Fermi level
in Fig.~\ref{fig:deltaPuBANDSa}(a)). As a next step we have 
used LDA+U+SO calculation scheme (\ref{U1}-\ref{SO-nondiag}) 
with non-spin-polarized electron density and LDA potential 
but keeping Hund parameter $J_H$ = 0.85~eV\cite{Savrasov00,SavrasovJ}
(see Fig.~\ref{fig:deltaPuBANDSa}(c)). In this calculation we have
found magnetic ground state ($S$ = 2.5, $L$ = 2.7, and $J$ = 0.2)
with a configuration close to $f^6$. And only the calculation by LDA+U+SO 
with Coulomb exchange parameter value obtained in constrain LDA
calculations ($J_H$ = 0.48~eV) gave (see Fig.~\ref{fig:deltaPuBANDSa}(d))
a nonmagnetic state with $S$ = 0, $L$ = 0, and $J$ = 0.

In the Sec.~\ref{Method} we have discussed that the use of 
spin-polarized potential based on LSDA is equivalent to neglecting
off-diagonal in spin variables exchange potential terms and hence enhances
the tendency to spin moment formation. To check this we have
performed LDA+U+SO calculations with the off-diagonal in spin
variables terms in LDA+U potential (Eq. \ref{Pot}) set to zero 
($U$ = 2.5~eV and $J_H$ = 0.48~eV). As a result we indeed obtained 
a magnetic ground state (see Table~\ref{tab:compcalc}) which is close 
to the results of LSDA+U calculations.\cite{Bouchet00,Savrasov01} 
To compare with the results of fully relativistic spin-polarized 
calculation we have performed LDA+U+SO calculations with 
the off-diagonal in spin variables terms in LDA+U potential 
(Eq. \ref{Pot}) set to zero using the value of Coulomb parameter 
$U$ = $J_H$ = 0.48~eV. In Ref.~\onlinecite{SolovyevUJ} it was shown 
that $U-J_H$ can be regarded as an effective Coulomb interaction 
parameter $U_{eff}$ in LSDA+U calculations, so the choice 
of $U$ = $J_H$ is equivalent to $U_{eff}$ = 0. Such calculations 
gave solution with a large values of spin and orbital moments 
(see Table~\ref{tab:compcalc}) similar to LSDA results.

\subsection{$\alpha$ phase of plutonium}
\label{Alpha-Pu}
\begin{figure}[!t]
\begin{center}
\epsfxsize=14cm
\epsfbox{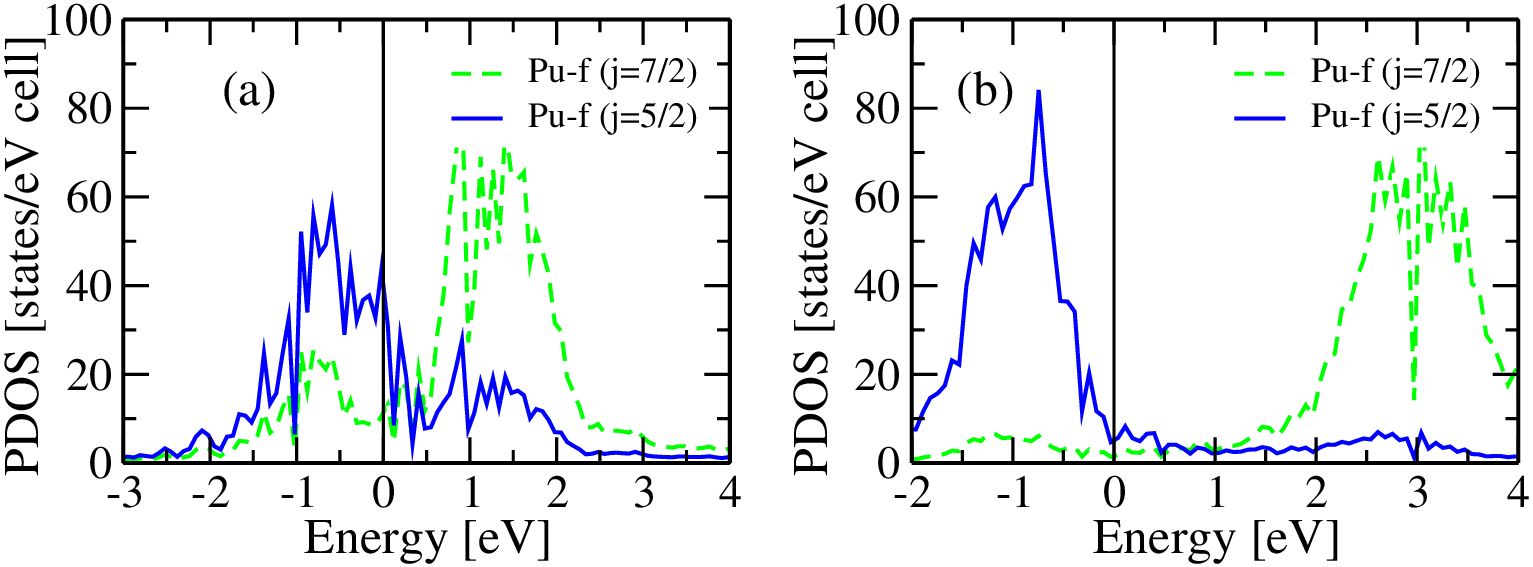}
\end{center}
\caption{Partial $f^{5/2}$ and $f^{7/2}$ contributions to 5$f$ band
densities of states of $\alpha$ plutonium calculated with LDA (a)
and LDA+U+SO method (b).}
\label{fig:alphaPuDOS}
\end{figure}
A transition from $\delta$ to $\alpha$ phase can be
described as a volume contraction on 19\% and a monoclinic
distortion of the fcc lattice. To separate the influence on the
electronic structure and magnetic properties of the volume
contraction and lattice distortion, we performed the LDA+U+SO
calculation for fcc lattice with the volume per Pu atom
corresponding to $\alpha$ phase. The result is the same
nonmagnetic solution: $S$ = 0, $L$ = 0, and $J$ = 0 in $f^6$ 
configuration as was obtained for $\delta$ phase. A variation 
of Hund parameter J$_H$ has shown that for Pu in fcc-structure 
with the volume of $\alpha$ phase nonmagnetic solution is 
a little more stable than for $\delta$ phase volume (while 
in the latter case already 15\% increase of $J_H$ was enough 
to produce the magnetic state, for $\alpha$ phase volume 
20\% increase is needed).

LDA+U+SO calculation for real monoclinic crystal structure of
$\alpha$ phase (see Fig.~\ref{fig:alphaPuDOS}) also gave a nonmagnetic
ground state with $S$ = 0, $L$ = 0, $J$ = 0 in $f^6$ configuration. 
Comparing with the results for $\delta$ phase (Fig.~\ref{fig:deltaPuDOS}) 
one can see that bands become more broad (Fig.~\ref{fig:alphaPuDOS}(a)) 
due to the smaller volume and hence increased
hybridization strength. Correspondingly $f^{5/2}$ and $f^{7/2}$
bands have a much more mixed character (a strong admixture of
$f^{5/2}$ orbitals to the nominally $f^{7/2}$ band and vice versa)
comparing with a more pure band character for $\delta$ phase
(Fig.~\ref{fig:deltaPuDOS}(a)). However, after taking into account
Coulomb interaction in LDA+U+SO calculations 
(Fig.~\ref{fig:alphaPuDOS}(b)) the band structure becomes very 
similar to the case of $\delta$ phase (Fig.~\ref{fig:deltaPuDOS}(b)). 
$f^{5/2}$ and $f^{7/2}$ bands have now pure orbital character.
$f^{5/2}$ band is nearly completely filled with the Fermi level
on the top of it and an $f^{7/2}$ band is empty. The separation
between the centers of those bands is $\approx$ 4~eV.

The calculated photoemission spectra (see Fig.~\ref{UPS}) is not
very different from the corresponding curve for the $\delta$ phase
except that it is more smooth and does not show high and lower
energy shoulders. The sharp peak near the Fermi level which is
stronger in experimental spectrum for the $\alpha$ phase than for
$\delta$ phase spectrum is missing in the calculated spectrum as
it was in the case for the $\delta$ phase. Again we expect that
Dynamical Mean-Field Theory\cite{KotliarVollhardt,DMFT,condmat}
(DMFT) (see Ref.~\onlinecite{Savrasov01,Savrasov04}) can solve
this problem. But the peak in experimental spectrum 
at $\approx$ 1~eV corresponding to the lower Hubbard band 
is well described by the calculated spectrum.

As we have shown above, the calculations scheme presented here is
different from the ones used in previous investigations of
plutonium electronic structure and magnetic properties. A
resulting nonmagnetic ground state (Pu ion in $f^6$
configuration with $jj$ coupling scheme) was never obtained in any
other calculations. In order to check validity of our method we
have performed calculations for a series of Pu compounds. The main
aim of this calculations was to investigate magnetic properties
for these compounds and compare the results with the available
experimental data.

\section{Plutonium compounds}
\label{Plutonium compounds}
\subsection{PuN}
\label{PuN}
\begin{figure}[!t]
\begin{center}
\epsfxsize=14cm
\epsfbox{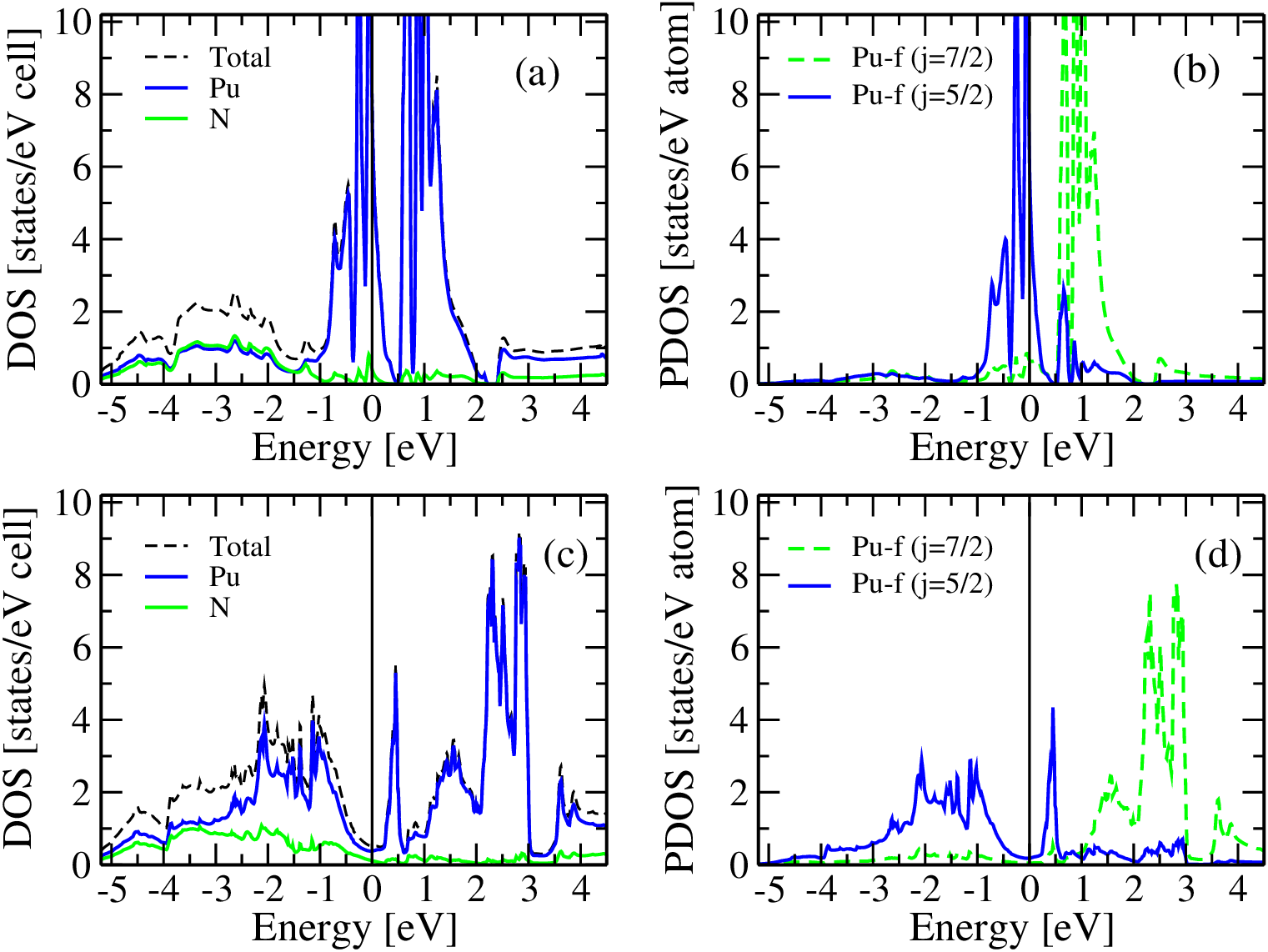}
\end{center}
\caption{(a) Total and partial densities of states of PuN calculated in LDA. (b) Partial $f^{5/2}$ and $f^{7/2}$ contributions
in Pu 5$f$ band for PuN from LDA calculations. (c) The same as (a) calculated by LDA+U+SO.
(d) The same as (b) calculated by LDA+U+SO.}
\label{fig:PuNDOS}
\end{figure}
\begin{table}
\caption{Electronic configuration of Pu ion in various plutonium compounds calculated
with LDA+U+SO method. Notations are the same as
for Table \ref{tab:deltaPua}.}
\begin{center}
\begin{tabular}{lllllllllll}
\hline \hline
Compound & OD$_{\{LS\}}$ & OD$_{\{jm_j\}}$ & largest eigenvalues & $S$ & $L$ & $J$ & $f^5$ & $f^6$ & $jj$ & $LS$\\
\hline
PuN         & 0.436 & 0.239 & 0.118 0.162 0.923 0.930 0.960 0.965 0.978 & 1.279 & 3.339 & 2.060 & 0.82 & 0.18 & 0.57 & 0.43\\
PuCoGa$_5$  & 0.426 & 0.226 & 0.029 0.796 0.883 0.897 0.922 0.939 0.941 & 1.144 & 1.594 & 0.450 & 0.18 & 0.82 & 0.62 & 0.38\\
PuRh$_2$    & 0.348 & 0.302 & 0.127 0.525 0.610 0.883 0.908 0.924 0.927 & 1.546 & 3.421 & 1.875 & 0.75 & 0.25 & 0.46 & 0.54\\
PuSi$_2$    & 0.441 & 0.142 & 0.070 0.907 0.916 0.939 0.949 0.960 0.962 & 0.692 & 0.829 & 0.137 & 0.05 & 0.95 & 0.77 & 0.23\\
PuTe        & 0.441 & 0.253 & 0.026 0.895 0.904 0.908 0.928 0.940 0.946 & 0.907 & 1.069 & 0.162 & 0.06 & 0.94 & 0.70 & 0.30\\
PuSb        & 0.453 & 0.320 & 0.046 0.127 0.959 0.961 0.973 0.981 0.982 & 1.583 & 3.650 & 2.067 & 0.83 & 0.17 & 0.44 & 0.56\\
\hline \hline
\end{tabular}
\end{center}
\label{tab:others}
\end{table}
\begin{table}
\caption{ Values of effective paramagnetic
moments (in Bohr's magnetons $\mu_B$) calculated from the total moment value $J$
 in $jj$ and $LS$ coupling schemes $\mu_{eff}^{LS}$ and $\mu_{eff}^{jj}$
 and their weighted value
$\mu_{eff}^{calc}$ in comparison with experimental data $\mu_{eff}^{exp}$ (see text).}
\begin{center}
\begin{tabular}{llllllllll}
\hline \hline
Compound & $J$& $jj$ & $LS$ & $\mu_{eff}^{jj}$ & %
$\mu_{eff}^{LS}$ & $\mu_{eff}^{calc}$ & $\mu_{eff}^{exp}$\\
\hline
PuN         & 2.060 & 0.57 & 0.43 & 2.16 & 0.72 & 1.54 &  1.08\cite{Raphael69}\\
PuCoGa$_5$  & 0.450 & 0.62 & 0.38 & 0.69 & 0.23 & 0.52 &  0.68\cite{Sarrao02}\\
PuRh$_2$    & 1.875 & 0.46 & 0.54 & 1.99 & 0.66 & 1.27 &  0.88\cite{Harvey73}\\
PuSi$_2$    & 0.137 & 0.77 & 0.23 & 0.34 & 0.11 & 0.23 &  0.54\cite{Olsen60,Boulet03}\\
PuTe        & 0.162 & 0.70 & 0.30 & 0.37 & 0.12 & 0.30 &  --\\
PuSb        & 2.067& 0.44 & 0.56 & 2.16 & 0.72 & 1.35 & 1.00\cite{Spirletunpub}\\
\hline \hline
\end{tabular}
\end{center}
\label{tab:others2}
\end{table}

Plutonium mononitride crystallizes in
a $rock$--$salt$-type structure with $a$ = 4.905~\AA.~\cite{PuNStr}
Neutron diffraction showed no long-range order and magnetic
moments larger than 0.25 $\mu_B$.~\cite{Boeuf84} From the magnetic
susceptibility and specific heat measurements an antiferromagnetic
transition at T$_N$ = 13~K was proposed.~\cite{Martin76} According
to another magnetic susceptibility curve, PuN is a Curie-Weiss
paramagnet with $\mu_{eff}$ = 1.08~$\mu_B$.~\cite{Raphael69}
Electronic structure of
PuN was calculated in L(S)DA with and without relativistic
correction.~\cite{Brooks84} SIC-LSDA method showed valency +5 to
be energetically preferred in the model with partially localized
5$f$ electrons of Pu ion.~\cite{Petit02}

Fig.~\ref{fig:PuNDOS}(a) and~\ref{fig:PuNDOS}(b) show total and
partial $f$ density of states of PuN obtained in our LDA with
spin-orbit coupling calculation (when the Coulomb interaction
correction was `switched off'). Nitrogen 2$p$-states are strongly
hybridized with Pu-$f$ states and the Fermi level is located
inside the $j$ = 5/2 subband. While $j$ = 5/2 and $j$ = 7/2
subbands are well separated from each other (see
~\ref{fig:PuNDOS}(b)), they do not have pure orbital character
with a strong admixture of $j$ = 5/2 states to the formally
$j$ = 7/2 subband. The results of LDA+U+SO calculation are presented
in Fig.~\ref{fig:PuNDOS}(c) and~\ref{fig:PuNDOS}(d). In contrast to
metallic Pu case Coulomb interaction correction not only has
increased energy separation between $j$ = 5/2 and $j$ = 7/2
subbands, but also led to the splitting of $j$ = 5/2 subband into
occupied and empty states (see the peak of $j$ = 5/2 character
just above the Fermi energy on~\ref{fig:PuNDOS}(d)). As this
split-off band contain one electron per Pu ion, one can conclude
that our results correspond to the configuration $f^5$ in
agreement with formal Pu valency +3 in PuN.

In Table~\ref{tab:others} (the first row corresponds to PuN) the
calculated values for spin, orbital and total moments are
presented. The total moment $J$ = 2.218 is close to an ideal value
$j$ = 5/2 for $f^5$ configuration. Five largest eigenvalues of the
occupation matrix also show a well defined $f^5$ configuration.
The analysis of occupation matrices off-diagonal elements show
that in $\{jm_j\}$ basis the values of those elements are nearly
two times smaller than the corresponding values for $\{LS\}$
basis. That corresponds to the intermediate coupling scheme closer
to $jj$ coupling. The analysis based on the calculated values for
spin, orbital, and total moments (see Eq. (\ref{LS-model}) in
Sec.~\ref{Delta-Pu}) gave 89\% $f^5$ and 11\% $f^6$
configuration with 57\% of $jj$ and 43\% of $LS$ couplings.

In order to compare our results with the experimental magnetic measurements
data we should obtain the magnetic moment value using
calculated values of spin ($S$), orbital ($L$), and
total moments ($J$). In the cases when spin-orbit coupling is weak and
Russell-Saunders coupling ($LS$ coupling) scheme is valid with
$\mathbf S$ and $\mathbf L$ operators well defined, this is a very simple
task: total magnetic moment value can be calculated as:
\begin{equation}
M_{tot}=2\cdot S+L
 \label{m=2s+l}
\end{equation}
However for strong spin-orbit coupling when $jj$ or intermediate coupling schemes
should be used, this problem becomes much more complicated. For the
intermediate coupling scheme there is no general solution of this problem
and for $jj$ coupling it can be solved only for the free ions in pure configuration.
An effective paramagnetic moment obtained from susceptibility measurements
using Curie-Weiss law can be calculated as:
\begin{equation}
\mu_{eff}=g\cdot\sqrt{J\cdot(J+1))}\cdot\mu_B
 \label{m-eff}
\end{equation}
The problem is to define Lande $g$-factor which can be calculated
for pure $f^5$ and $f^6$ configurations in $LS$ or $jj$ coupling
schemes. As for $f^6$ configuration total moment value $J$ = 0,
one need to calculate $g$-factor for $f^5$ configuration only. For
ground state of $f^5$ configuration in $jj$ coupling scheme Lande
factor $g_{jj}$ = 6/7 $\approx$ 0.86. In $LS$ coupling scheme its
value is $g_{LS}$ = 2/7 $\approx$ 0.29. As the latter value is
nearly three times larger than the former, $g_{jj}$ and $g_{LS}$
can give only an upper and lower limits of $g$-factor for the case
of intermediate coupling.

In Table~\ref{tab:others2} the values of effective paramagnetic
moments calculated using Eq. (\ref{m-eff}) are presented. In addition to
$\mu_{eff}^{LS}$ and $\mu_{eff}^{jj}$ calculated using Lande factors $g_{LS}$
and $g_{jj}$ correspondingly, we have calculated their weighted value
$\mu_{eff}^{calc}$ using relative weight of $LS$ and $jj$ coupling
obtained from Eq. (\ref{LS-model}).

Results for PuN (the first row of the Table \ref{tab:others2})
gave $\mu_{eff}^{calc}$ = 1.54~$\mu_B$ in a reasonable agreement
with the experimental value $\mu_{eff}^{exp}$ = 1.08~$\mu_B$.

\subsection{PuCoGa$_5$}
\label{PuCoGa5}
\begin{figure}[!t]
\begin{center}
\epsfxsize=14cm
\epsfbox{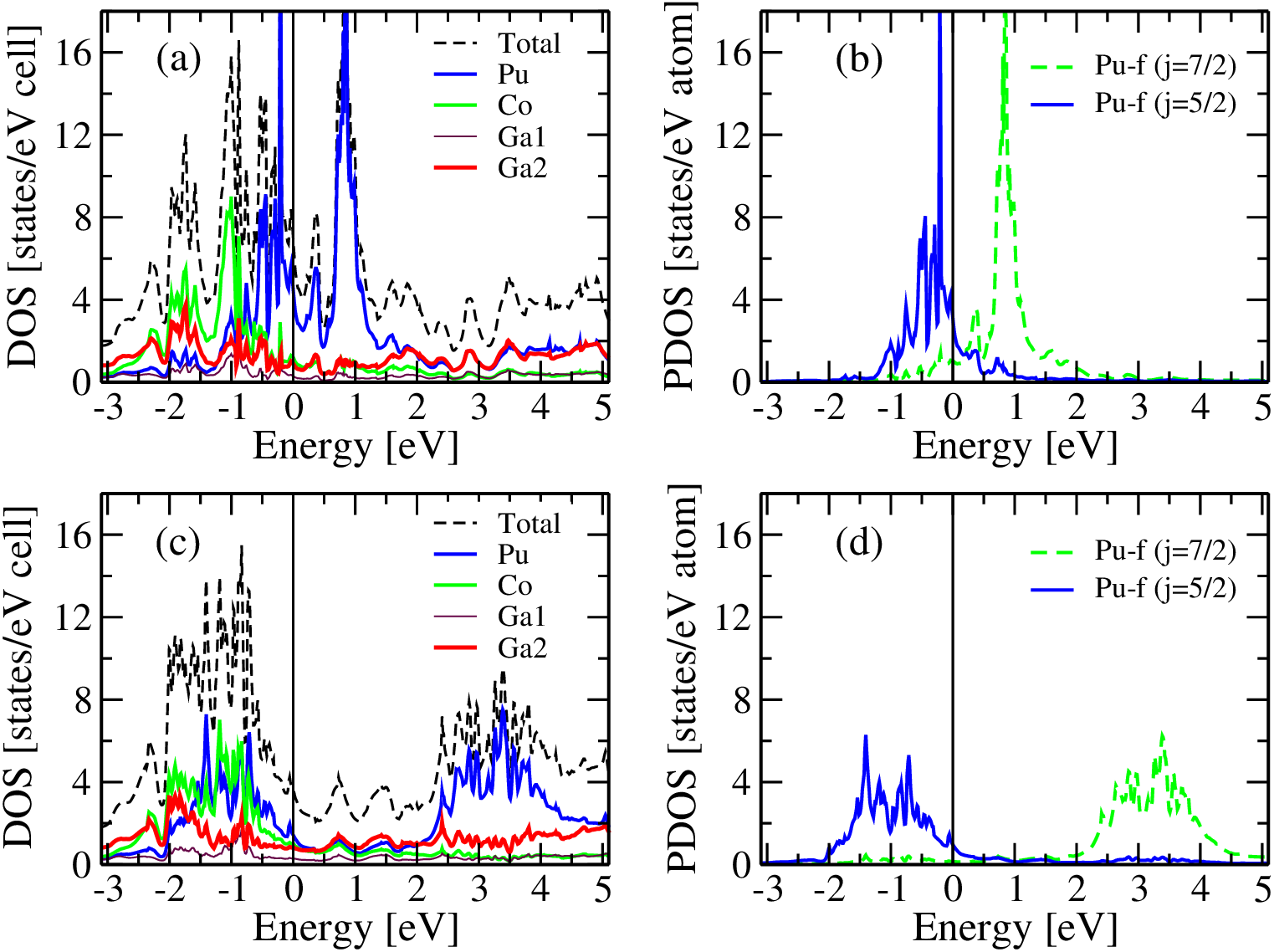}
\end{center}
\hspace{-3 cm}
\caption{(a) Total and partial densities of states of PuCoGa$_5$ calculated in LDA. 
(b) Partial $f^{5/2}$ and $f^{7/2}$ contributions in Pu 5$f$ band for PuCoGa$_5$ 
from LDA calculations. (c) The same as (a) calculated by LDA+U+SO. (d) The same 
as (b) calculated by LDA+U+SO.}
\label{fig:PuCoGa5}
\end{figure}
Since first reported in 2002 by Sarrao $et~al.$,~\cite{Sarrao02}
superconductor PuCoGa$_5$ has attracted a special attention being
the first Pu-containing superconductor in well-known
superconductor class of `115' materials.~\cite{Opahle04,115a,115b} Its
superconducting temperature $T_c$ = 18.5~K is an order of
magnitude higher than for other `115' superconductors.~\cite{115a,115b}
Specific heat coefficient value $\gamma$ = 77 mJ mol$^{-1}$
K$^{-2}$ above T$_C$ and Curie-Weiss behavior of magnetic
susceptibility with $\mu_{eff}$ = 0.68 $\mu_B$ are indicative of
unconventional superconductivity of
PuCoGa$_5$.~\cite{Sarrao02,Sarrao03,Thompson04,Griveau04,Bang04,Bauer04}
Tanaka {\it et al.}\cite{Tanaka04} used periodic Anderson model to
describe $d$-wave superconducting state of PuCoGa$_5$. Its
isostructural counterpart PuRhGa$_5$ shows analogous properties
but with the lower value of SC transition temperature
$T_c$ = 8 K.~\cite{Wastin03}

PuCoGa$_5$ crystallizes in tetragonal $P4/mmm$ space group with
$a$ = 4.232 \AA~and $c$ = 6.782 \AA.~\cite{Sarrao02} Electronic
structure calculation of PuCoGa$_5$ was performed in fully
relativistic full-potential method of local
orbitals.~\cite{Opahle03} In that work the paramagnetic state has
a total energy value substantially higher than FM and AFM
solutions.

The LDA calculation without spin-orbit coupling\cite{Opahle04,Szajek03} 
showed the Fermi level crossing Pu 5$f$ bands manifold without 
any splitting. Another calculation by relativistic LAPW 
method\cite{Maehira03} gave Fermi surface analogous 
to Ce$M$In$_5$ series. Microscopic model for compounds 
with $f$ ions in $jj$ coupled state was extended 
to PuCoGa$_5$ in Ref.~\onlinecite{Hotta03}. Antiferromagnetic order 
was proposed in electronic structure calculation 
in Ref.~\onlinecite{Soderlind04a}. For our knowledge, in this work 
we report the first electronic structure calculation for PuCoGa$_5$ 
with Coulomb interactions taken into account beyond LDA or GGA 
approximations.

In PuN only $5f$-bands are present at the Fermi level while fully
occupied N-$p$ band is situated substantially lower
(Fig.~\ref{fig:PuNDOS}(a)). In contrast to that in PuCoGa$_5$
Co-$d$ and Ga-$p$ bands cross the Fermi level in addition to 5$f$
bands of Pu (see Fig.~\ref{fig:PuCoGa5}(a)) so that 5$f$ states
contribute only one third to the of density of states value at the
Fermi energy. This fact leads to a very complicated general band
structure for this compound (Fig.~\ref{fig:PuNDOS}(a)) but the
partial 5$f$ densities of state presented on
Fig.~\ref{fig:PuNDOS}(b) shows well pronounced $f^{5/2}$ and
$f^{7/2}$ subbands with $\approx$ 1.5~eV separation between them.
This picture is very close to the results for $\delta$ phase of
metallic Pu (Fig.~\ref{fig:deltaPuDOS}(b)) but with a larger
overlapping of $f^{5/2}$ and $f^{7/2}$ subbands and the position
of Fermi level a little deeper inside the $f^{5/2}$ band. This 
is an effect of a stronger hybridization of 5$f$ orbitals with 
Co-$d$ and Ga-$p$ states in PuCoGa$_5$ comparing with metallic Pu.

The Coulomb interaction correction in LDA+U+SO
calculations (Fig.~\ref{fig:PuNDOS}(c) and (d)) results in a larger energy
separation between $f^{5/2}$ and $f^{7/2}$ subbands (till 4~eV)
and nearly pure orbital character of these bands, again very
similar to metallic Pu (Fig.~\ref{fig:deltaPuDOS}(d)). In
Table~\ref{tab:others} (the second row corresponds to PuCoGa$_5$)
the calculated values for spin, orbital and total moments are
presented. Pu ion for our solution is predominantly in $f^6$
configuration but with a significant (18\%) admixture of $f^5$
configuration. Sizable values of spin and orbital moment together
with relatively small value of total moment gave similar to PuN
case an intermediate coupling scheme closer to $jj$ coupling.

We would like to note, that in spite of very similar to metallic
Pu band structure for 5$f$ states (compare Fig.~\ref{fig:PuCoGa5}(b) 
and Fig.~\ref{fig:deltaPuDOS}(b)) in the LDA
calculations, LDA+U+SO calculations scheme gave a magnetic ground
state for PuCoGa$_5$ in agreement with experimental
data.~\cite{Sarrao02,Sarrao03,Thompson04,Griveau04,Bang04}
In Table~\ref{tab:others2} (second row) the values of effective 
paramagnetic moments calculated using Eq. (\ref{m-eff}) are presented. 
The weighted value $\mu_{eff}^{calc}$ = 0.52~$\mu_B$ is smaller 
than experimental $\mu_{eff}^{exp}$ = 0.68~$\mu_B$, which is very 
close to the $\mu_{eff}^{jj}$ = 0.69~$\mu_B$ calculated with Lande 
$g$-factor $g_{jj}$ obtained in $jj$ coupling scheme.
\begin{figure}[!t]
\begin{center}
\epsfxsize=14cm
\epsfbox{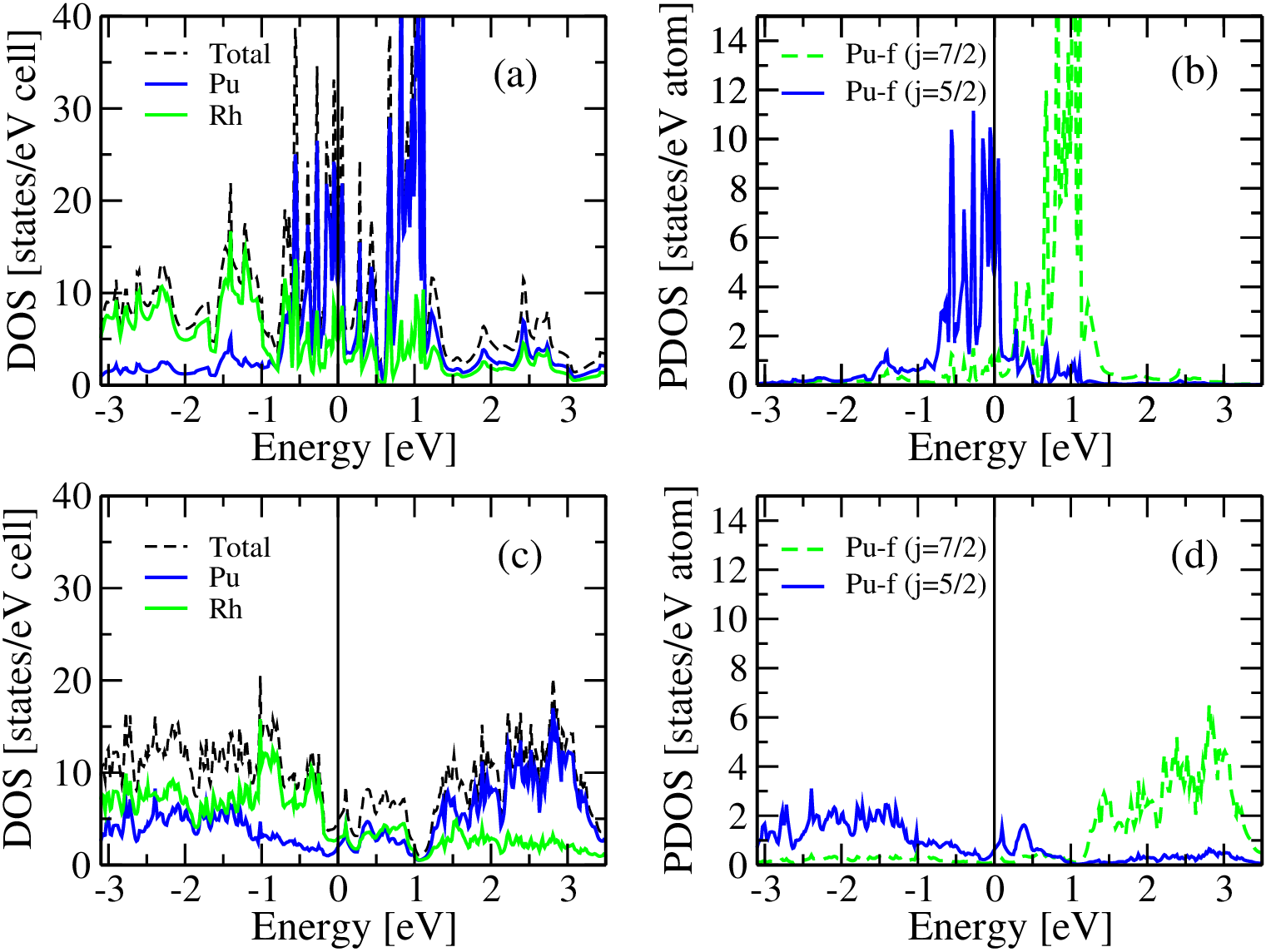}
\end{center}
\hspace{-3 cm}
\caption{(a) Total and partial densities of states of PuRh$_2$ calculated in LDA. (b) Partial $f^{5/2}$ and $f^{7/2}$ contributions
in Pu 5$f$ band for PuRh$_2$ from LDA calculations. (c) The same as (a) calculated by LDA+U+SO.
(d) The same as (b) calculated by LDA+U+SO.}
\label{fig:PuRh2}
\end{figure}

\subsection{PuRh$_2$}
\label{PuRh2}
PuRh$_2$ crystallizes in Laves structure (C15) $a$ = 7.488
\AA.~\cite{Harvey73} From the magnetic
susceptibility\cite{Harvey73} PuRh$_2$ is a Curie-Weiss
paramagnet with $p_{eff}$ = 0.88~$\mu_B$, $\Theta_{p}$ = --49~K,
and $\chi_{max}$ = 4.4 $\times$ 10$^{-3}$~emu mol$^{-1}$. Resistivity
agrees well with the magnetic susceptibility
results.~\cite{Harvey73} According to the specific heat results
with $\gamma$ = 145~mJ mol$^{-1}$ K$^{-2}$ and $\theta_D$ = 190~K,
PuRh$_2$ was classified as `middle-weight' fermion system without
temperature dependence of $\gamma$ in the region of low
temperatures.~\cite{Stewart84}

The dominant contribution to the PuRh$_2$ band structure gives a
broad partially filled Rh-$d$ band (Fig.~\ref{fig:PuRh2}(a)). Pu
5$f$ states (Fig.~\ref{fig:PuRh2}(b)) show significant
hybridization with Rh-$d$ band. However the general feature common
to metallic Pu and all Pu compounds investigated in this work, a
separation into $f^{5/2}$ and $f^{7/2}$ subbands still can be seen
here (Fig.~\ref{fig:PuRh2}(b)). The position of Fermi level inside
the $f^{5/2}$ band similar to PuN case (Fig.~\ref{fig:PuNDOS}(b))
shows that resulting configuration should be close to $f^5$.
Indeed LDA+U+SO calculations gave (Fig.~\ref{fig:PuRh2}(c) and
(d)) the $f^{5/2}$ states split into occupied and empty bands. In
contrast to PuN case (Fig.~\ref{fig:PuNDOS}(d)) empty $f^{5/2}$
band is not narrow but rather broad and has a two peak structure
due to the strong hybridization with Rh-$d$ states.

In Table~\ref{tab:others} (the third row corresponds to PuRh$_2$)
the calculated values for spin, orbital and total moments are
presented. In agreement with the aforementioned density of states
analysis, the Pu ion is predominantly in $f^5$ configuration with
a significant admixture (25\%) of $f^6$ configuration. A large
values of spin moment results in intermediate coupling scheme
closer to $LS$ coupling. In Table~\ref{tab:others2} (third row) 
the values of effective paramagnetic moments calculated using 
Eq. (\ref{m-eff}) are presented. The weighted 
value $\mu_{eff}^{calc}$ = 1.27~$\mu_B$ is larger than
experimental $\mu_{eff}^{exp}$ = 0.88~$\mu_B$ which still lies in
the limits of $\mu_{eff}^{jj}$ = 1.99~$\mu_B$ and $\mu_{eff}^{LS}$
= 0.66~$\mu_B$ closer to $LS$ coupling value $\mu_{eff}^{LS}$.
\begin{figure}
\begin{center}
\epsfxsize=14cm
\epsfbox{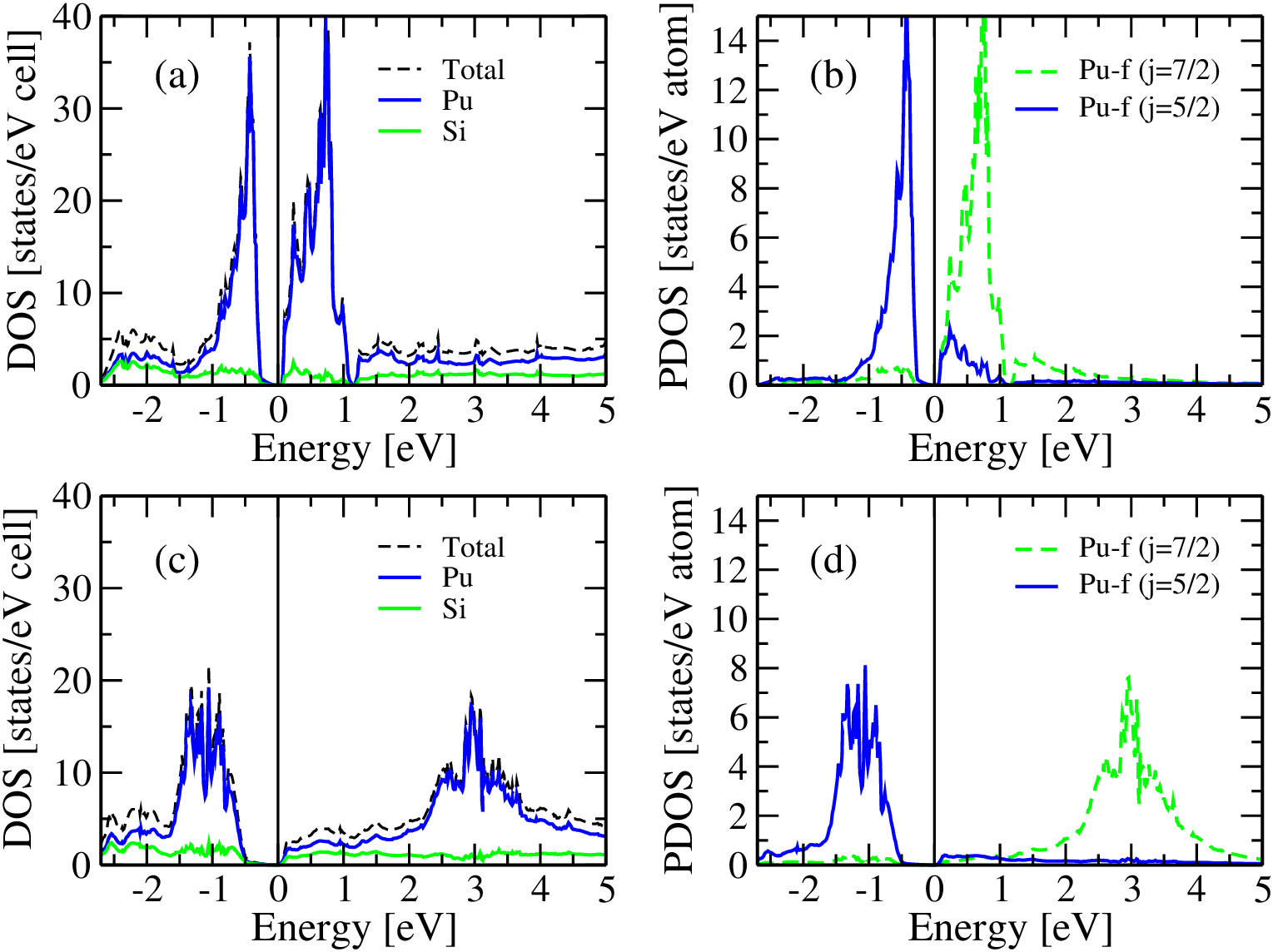}
\end{center}
\hspace{-3 cm}
\caption{(a) Total and partial densities of states of PuSi$_2$ calculated in LDA. 
(b) Partial $f^{5/2}$ and $f^{7/2}$ contributions in Pu 5$f$ band for PuSi$_2$ 
from LDA calculations. (c) The same as (a) calculated by LDA+U+SO. (d) The same 
as (b) calculated by LDA+U+SO.}
\label{fig:PuSi2}
\end{figure}

\subsection{PuSi$_2$}
\label{PuSi2} Structural and magnetic properties of PuSi$_2$ were
reported by Boulet $et~al.$~\cite{Boulet03} although this phase
was earlier investigated in Ref.~\onlinecite{Zachariasen49}. This
compound crystallizes in tetragonal $ThSi_2$-type (space group $I4_1/amd$)
structure with $a$ = 3.9707(3)~\AA~and $c$ = 13.6809(5)~\AA.~\cite{Boulet03} 
PuSi$_2$ susceptibility curve shows Curie-Weiss behavior\cite{Olsen60,Boulet03} 
with $p_{eff}$ = 0.54~$\mu_B$, $\Theta_p$ = --58~K, and $\chi_0$ = 2.3 $\times$ 
10$^{-5}$~emu mol$^{-1}$ (Ref.~\onlinecite{Boulet03}). Almost field independent 
resistivity shows a broad maximum at 18~K suggesting strong spin 
fluctuations.~\cite{Boulet03}

The LDA band structure (Fig.~\ref{fig:PuSi2}(a)) shows a broad deep
pseudogap with a very small density of states value on the Fermi
level. This pseudogap separates $f^{5/2}$ and $f^{7/2}$ subbands
(Fig.~\ref{fig:PuSi2}(b)) and position of the Fermi level exactly
in the pseudogap gave a completely filled $f^{5/2}$ band and hence
a pure $f^6$ configuration could be expected. Indeed LDA+U+SO
calculations gave (Fig.~\ref{fig:PuSi2}(c) and (d)) a solution
with the same type of band structure as without Coulomb
interaction correction except increased energy separation between
$f^{5/2}$ and $f^{7/2}$ subbands.

In Table~\ref{tab:others} (the fourth row corresponds to PuSi$_2$)
the calculated values for spin, orbital and total moments are
presented. Pu ion is in nearly pure $f^6$ configuration with a
small (5\%) contribution of $f^5$ configuration. Relatively small
values of spin and orbital moments gave an intermediate coupling
scheme very close to pure $jj$ coupling. The weighted value
of effective paramagnetic moment $\mu_{eff}^{calc}$ = 0.23~$\mu_B$
(Table~\ref{tab:others2}) is smaller than experimental
$\mu_{eff}^{exp}$ = 0.54~$\mu_B$.
\begin{figure}
\begin{center}
\epsfxsize=14cm
\epsfbox{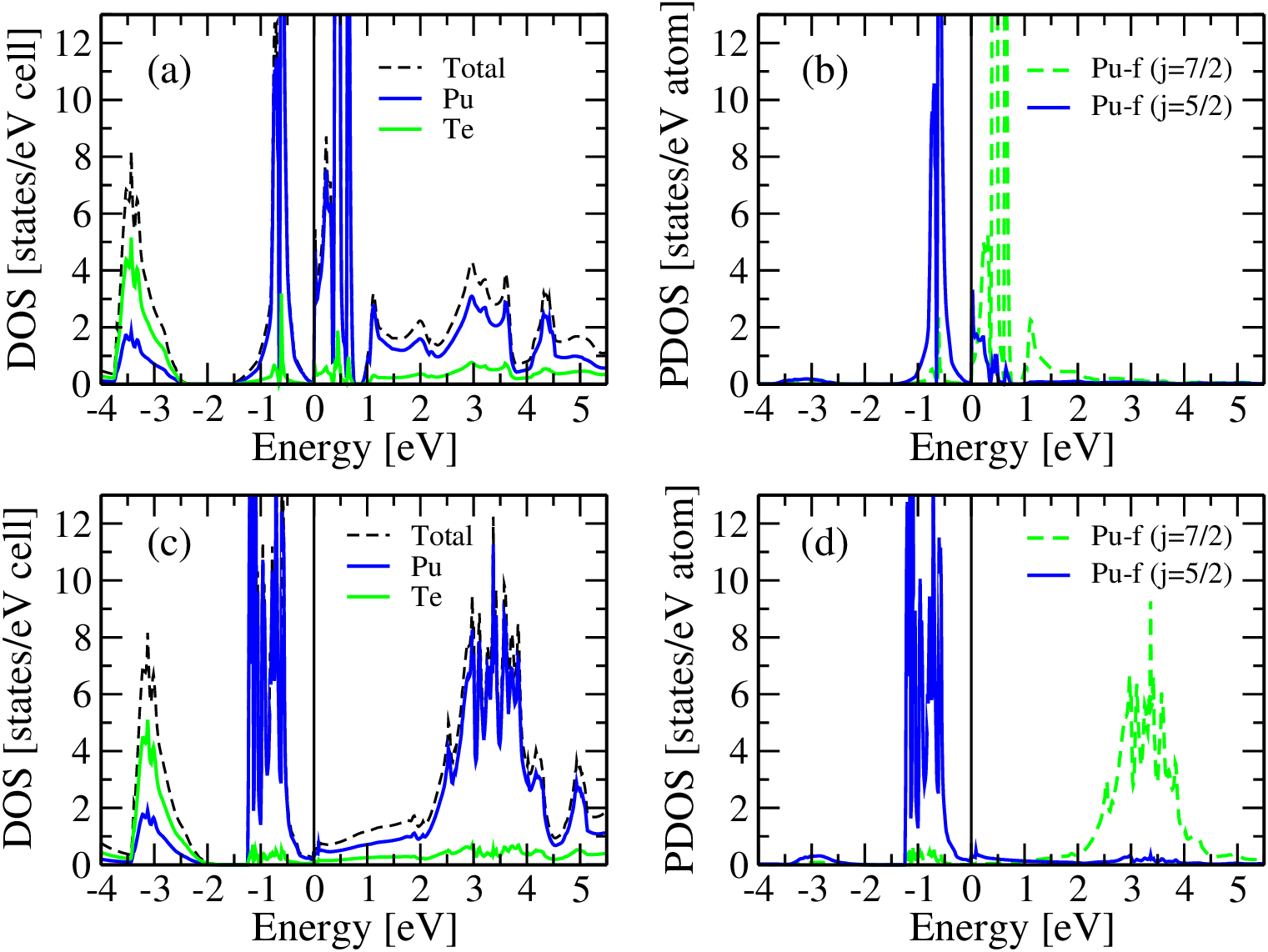}
\end{center}
\hspace{-3 cm}
\caption{(a) Total and partial densities of states of PuTe calculated in LDA. (b) Partial $f^{5/2}$ and $f^{7/2}$ contributions
in Pu 5$f$ band for PuTe from LDA calculations. (c) The same as (a) calculated by LDA+U+SO.
(d) The same as (b) calculated by LDA+U+SO.}
\label{fig:PuTe}
\end{figure}

\subsection{PuTe}
\label{PuTe}
PuTe crystallizes in $NaCl$-type structure with $a$ = 6.183~\AA.
Resistivity shows narrow-gap semiconductor behavior and
an electronic specific heat coefficient has a high value
$\gamma$ = 60~mJ mol$^{-1}$ K$^{-2}$ (Ref.~\onlinecite{Fournier90}).
In another work~\cite{Stewart91} the value of $\gamma$ = 30~mJ mol$^{-1}$
K$^{-2}$ was reported. While formal valency of Pu in PuTe is equal
+2, an intermediate valent Pu$^{+2}$-Pu$^{+3}$ state in this compound
was proposed.~\cite{Lander88,Wachter91,Mendik93,Wachter03}
Magnetic and optical properties of PuTe show a number of
peculiarities.~\cite{Burlet89,Olsen92,Abraham96}
A structural phase transition from NaCl to CsCl phase was proposed
from resistivity measurements.~\cite{Ichas01}

Electronic structure of PuTe was calculated by Dirac
equation-corrected ASA,~\cite{Brooks87} relativistic
LAPW,~\cite{Hasegawa92} and LSDA augmented plane wave method
(ASW).~\cite{Oppeneer00} From these calculation, $f^{7/2}$ and
$f^{5/2}$ subbands are split on 1~eV\cite{Brooks87} (or 0.3~eV
in Ref.~\onlinecite{Hasegawa92}). Results strongly depend on spin-orbit 
coupling strength.~\cite{Oppeneer00} Based on the M\"ossbauer 
spectra results,~\cite{Sanchez87} magnetic transition going from 
PuSb to PuTe with the vanishing of local moments was suggested.

Results of calculations by LDA and LDA+U+SO are presented in
Fig.~\ref{fig:PuTe} and in Table~\ref{tab:others} (the fifth row
corresponds to PuTe). The electronic structure and magnetic state
are very similar to those for PuSi$_2$, only the pseudogap is not so well
developed. Nearly pure $f^6$ configuration corresponds to the
formal Pu valency +2 with a small values of magnetic moments
described by 30\% admixture of $LS$ coupling to $jj$ coupling
scheme. The arguments against valency +2 in
Ref.~\onlinecite{Lander88,Wachter91,Mendik93,Wachter03} were based on
the very large ionic radius value for Pu$^{+2}$ ion. However, in spite
of the formal configuration $f^6$ in our results, the total number
of 5$f$ electrons is equal to 5.65 (effect of strong hybridization
of 5$f$orbitals with Sb-p states), which corresponds to the
intermediate valency Pu$^{+2}$-Pu$^{+3}$. Another interesting fact
is that going from PuTe to PuSb the number of 5$f$ electrons
decreases from 5.65 to 5.16, two times smaller than
expected from formal valency difference. A very small value of
magnetic moments obtained in our calculations (Table \ref{tab:others2})
agrees well with nonmagnetic state of Pu ions in PuTe.~\cite{Sanchez87}
\begin{figure}
\begin{center}
\epsfxsize=14cm
\epsfbox{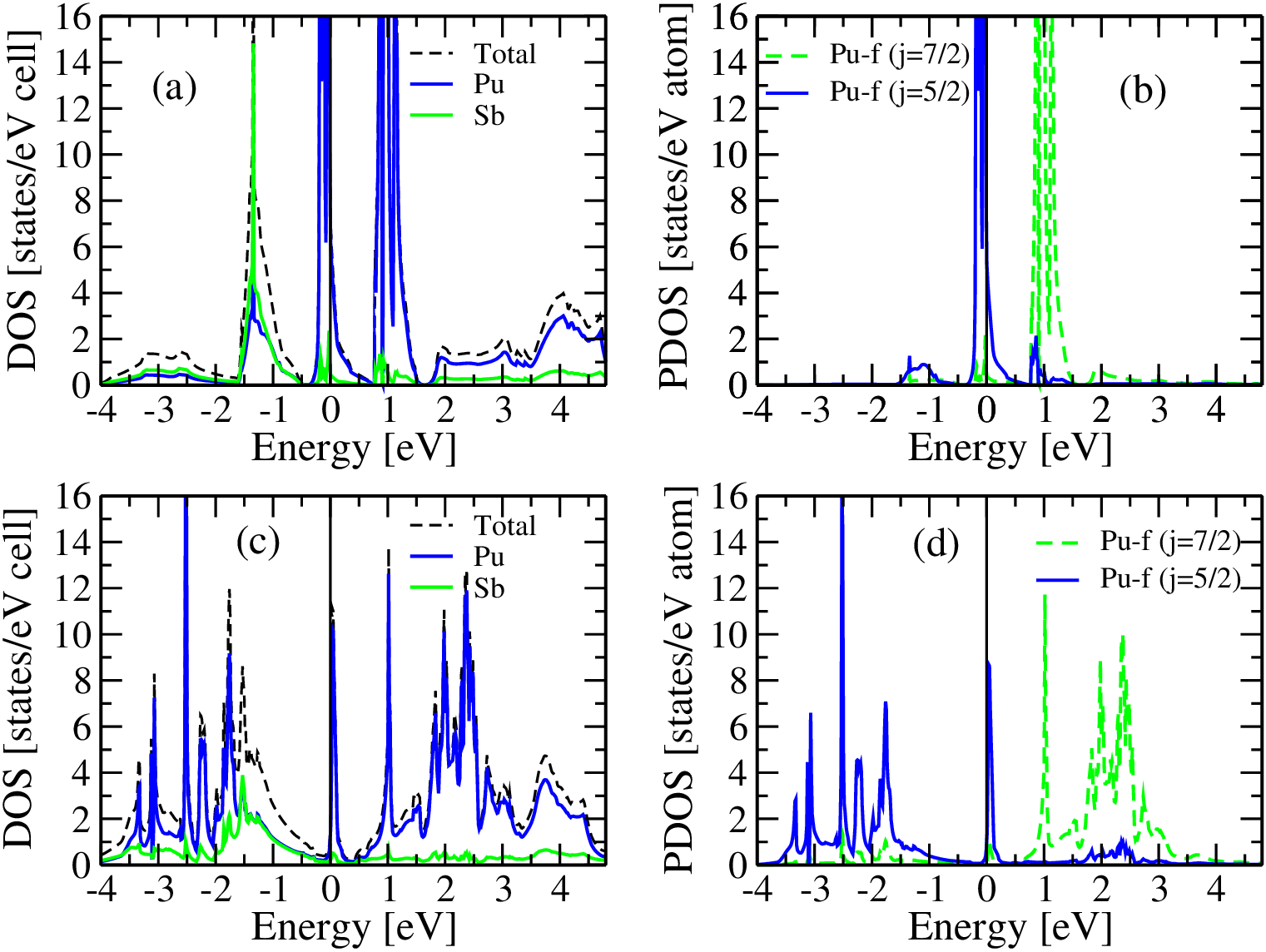}
\end{center}
\hspace{-3 cm}
\caption{(a) Total and partial densities of states of PuSb calculated in LDA. 
(b) Partial $f^{5/2}$ and $f^{7/2}$ contributions in Pu 5$f$ band for PuSb 
from LDA calculations. (c) The same as (a) calculated by LDA+U+SO. (d) The same 
as (b) calculated by LDA+U+SO.}
\label{fig:PuSb}
\end{figure}

\subsection{PuSb}
\label{PuSb} From magnetization measurements and neutron
experiments an antiferromagnetic spin structure was found with a
strong $<$001$>$ anisotropy and ordering temperature 75
K.~\cite{Cooper83,Lander84} Preferable intermediate coupling
type closer to $LS$-type was proposed in Ref.~\onlinecite{Burlet84} 
(model of wave function consisting of 90\% $LS$- and 10 \% $jj$-coupled 
wave functions\cite{Cooper83} gave a possibility to AFM transition leaving 
strong anisotropy). Neutron-diffraction experiment reveals low-temperature 
Pu ion magnetic moment $\mu$ = 0.75~$\mu_B$ perpendicular to ferromagnetic 
(001) planes with N\'eel temperature T$_N$ = 85~K and first-order transition 
to the incommensurate ferromagnetic
phase.~\cite{Burlet84,Lander86,Burlet89} Effective magnetic moment
of paramagnetic phase from susceptibility was estimated as
$\mu$ = 1.0~$\mu_B$.~\cite{Spirletunpub} Model of localized 5$f$ electrons
with weak hybridization to conducting electron states
was proposed.~\cite{Lander86,Hu87} Electrical resistivity
shows metallic behavior with Kondo-like broad maximum at
106 K.~\cite{Blaise85} X-ray photoelectron spectroscopy (XPS) and
high-resolution valence-band ultraviolet photoelectron
spectroscopy (UPS) showed strong localized nature of 5$f$ Pu
electrons in PuSb and pronounced $f^5$ configuration.~\cite{Gouder00}

Results of calculations by LDA and LDA+U+SO are presented in
Fig.~\ref{fig:PuSb} and in Table~\ref{tab:others} (the sixth row
corresponds to PuSb). The electronic structure and magnetic state
are very similar to PuN, according to the formal valency +2 for Pu
in both compounds.

We have calculated photoemission spectrum for PuSb (see Fig.~\ref{UPS}).
An agreement with experimental curve\cite{Durakiewicz04} is a good one.
The main peak at $\approx$ 1.5~eV corresponding to Pu-5$f$ states
(see Fig.~\ref{fig:PuSb}(c)) is reproduced quite well.

In Table~\ref{tab:others2} (sixth row corresponds to PuSb) the
values of effective paramagnetic moments calculated using Eq. (\ref{m-eff}) 
are presented. The calculated $\mu^{calc}_{eff}$ =
1.35~$\mu_B$ is close to experimental value $\mu_{eff}^{exp}$ =
1.00~$\mu_B$.

\section{Conclusion}
\label{Dis}

In this work we present the results of theoretical investigation for
Pu ion magnetic state in metallic plutonium and plutonium
compounds. In contrast to all previous theoretical studies
but in agreement with experimental measurements we have found
for metallic Pu in both $\alpha$ and $\delta$ phases
a nonmagnetic ground state with $f^6$ configuration of
5$f$ shell in pure $jj$ coupling scheme. A strong spin-orbit coupling
for 5$f$ electrons results in a splitting of $f$ bands into well separated
$f^{5/2}$ and $f^{7/2}$ subbands with $f^{5/2}$ band nearly fully
occupied and $f^{7/2}$ band empty giving `preformed' $f^6$ configuration.
Taking into account Coulomb interaction via LDA+U potential
does not change this nonmagnetic ground state.

We have shown that approximately equal strength of spin-orbit
coupling and exchange interaction, whose matrices can not be made
simultaneously diagonal in the same basis set, does not allow the use of 
the simplified diagonal forms of the corresponding Hamiltonian terms.
That is true also for spin-polarized potential used in LSDA. Only
a general non-diagonal matrix form of exchange interaction is 
appropriate for 5$f$ electrons.

We have calculated also a series of plutonium compounds with different
formal valency and the calculated magnetic moments values agree
reasonably well with experimental data.

Comparison of the calculated and experimental photoemission spectra
show that while LDA+U+SO method based on static mean-field
approximation can reproduce Hubbard bands appearing in such
strongly correlated materials as plutonium and its compounds, the quasiparticle
peak on the Fermi energy can not be described by this method.
The more elaborate Dynamical Mean-Field Theory (DMFT) is needed here.
The results obtained in LDA+U+SO calculations could be used
a basis for further DMFT studies.

\section{Acknowledgments}
The authors are indebted to A.~I.~Poteryaev, J.~E.~Medvedeva, and
A.~V.~Kozhevnikov for contributing to the development of the
LDA+U+SO code. This work was partly supported by the Russian
Foundation for Basic Research grant RFFI-04-02-16096.


\begin{thebibliography}{99}
\bibitem{Freeman74}
{\it The Actinides: Electronic Structure and Related Properties},
edited by A.~J.~Freeman and J.~B.~Darby, Jr. (Academic Press, New York, 1974),
Vols. I and II.
\bibitem{HeckerPMS}
S.~S.~Hecker, D.~R.~Harbur, and T.~G.~Zocco,
Prog. Mater. Sci. {\bf 49}, 429 (2004).
\bibitem{Lander03}
G.~H.~Lander,
Science {\bf 301}, 1057 (2003).
\bibitem{LosAlamos}
{\it Challenges in Plutonium Science},
edited by N.~G.~Cooper, Los Alamos Sci. {\bf 26} (LANL, Los Alamos, NM, 2000).
\bibitem{MRS}
Current state of Pu problem is reviewed
in Mat. Res. Soc. Bull. {\bf 26}, No. 9 (September), 2001;
e.g., S.~S.~Hecker, {\it ibid}. {\bf 26}, 672 (2001).
\bibitem{Albers01}
R.~C.~Albers,
Nature {\bf 410}, 759 (2001).
\bibitem{Moore03}
K.~T.~Moore, M.~A.~Wall, A.~J.~Schwartz, B.~W.~Chung, D.~K.~Shuh,
R.~K.~Schulze, and J.~G.~Tobin,
Phys. Rev. Lett. {\bf 90}, 196404 (2003);
K. T. Moore, M. A. Wall, A. J. Schwartz, B. W. Chung, S. A. Morton, J. G. Tobin, S. Lazar,
F. D. Tichelaar, H. W. Zandbergen, P. S\"oderlind, and G. van der Laan,
Philos. Mag. {\bf 84}, 1039 (2004).
\bibitem{Laan04}
G.~van~der~Laan, K.~T.~Moore, J.~G.~Tobin, B.~W.~Chung, M.~A.~Wall, and A.~J.~Schwartz,
Phys. Rev. Lett. {\bf 93}, 097401 (2004);
G.~var~der~Laan and B.~T.~Thole,
Phys. Rev. B {\bf 53}, 14458 (1996).
\bibitem{Cooper85}
B.~R.~Cooper, R.~Siemann, D.~Yang, P.~Thayamballi, and A.~Banerjea,
in {\it Handbook on the Physics and Chemistry of the Actinides}, edited by
A.~J.~Freeman and G.~H.~Lander (Elsevier, Amsterdam, 1985), Vol. 6, p. 435.
\bibitem{PESa}
A.~J.~Arko, J.~J.~Joyce, L.~A.~Morales, J.~H.~Terry, and R.~K.~Schulze,
Los Alamos Sci. {\bf 26}, 168 (2000).
\bibitem{PESb}
A.~J.~Arko, J.~J.~Joyce, L.~Morales, J.~Wills, J.~Lashley, F.~Wastin, and J.~Rebizant,
Phys. Rev. B {\bf 62}, 1773 (2000);
J.~Terry, R.~K.~Schulze, J.~D.~Farr, T.~Zocco, K.~Heinzelman,
E.~Rotenberg, D.~K.~Shuh, G.~van~der~Laan, D.~A.~Arena, and J.~G.~Tobin,
Surf. Sci. {\bf 499}, L141 (2002).
\bibitem{Naegele86}
J.~R.~Naegele, L.~Manes, J.~C.~Spirlet, and W.~M\"uller,
Phys. Rev. Lett. {\bf 52}, 1834 (1986);
J.~R.~Naegele, J.~Ghijsen, and L.~Manes,
{\it Actinides -- Chemistry and Physical Properties, Structure and Bounding 59/60}
(Springer-Verlag, Berlin, 1985), p. 197.
\bibitem{Havela02}
L.~Havela, T.~Gouder, F.~Wastin, and J.~Rebizant,
Phys. Rev. B {\bf 65}, 235118 (2002).
\bibitem{Tobin03}
J.~G.~Tobin, B.~W.~Chung, R.~K.~Schulze, J.~Terry, J.~D.~Farr, D.~K.~Shuh, K.~Heinzelman,
E.~Rotenberg, G.~D.~Waddill, and G.~van~der~Laan,
Phys. Rev. B {\bf 68}, 155109 (2003).
\bibitem{Wick80}
{\it Plutonium Handbook: a Guide to the Technology}, edited by O.~J.~Wick (American
Nuclear Society, LaGrange Park, IL, 1980).
\bibitem{Lashley03}
J.~C.~Lashley, J.~Singleton, A.~Migliori, J.~B.~Betts, R.~A.~Fisher, J.~L.~Smith,
and R.~J.~McQueeney,
Phys. Rev. Lett. {\bf 91}, 205901 (2003).
\bibitem{Wong03}
J.~Wong, M.~Krisch, D.~L.~Farber, F.~Occelli, A.~J.~Schwartz, T.-C.~Chiang, M.~Wall,
C.~Boro, and R.~Xu,
Science {\bf 301}, 1078 (2003).
\bibitem{Ledbetter76}
H.~M.~Ledbetter and R.~L.~Moment,
Acta Metall. {\bf 24}, 891 (1976).
\bibitem{McQueeney04}
Recently, this effect of softening was found from inelastic neutron scattering on
polycrystals of $\delta$ phase $^{242}$Pu$_{0.95}$Al$_{0.05}$ by another experimental group:
R.~J.~McQueeney, A.~C.~Lawson, A.~Migliori, T.~M.~Kelley, B.~Filtz, M.~Ramos, B.~Martinez,
J.~C.~Lashley, and S.~V.~Vogel,
Phys. Rev. Lett. {\bf 92}, 146401 (2004).
\bibitem{Dai03}
X.~Dai, S.~Y.~Savrasov, G.~Kotliar, A.~Migliori, H.~Ledbetter, and E.~Abrahams,
Science {\bf 300}, 953 (2003).
\bibitem{Lashleycondmat}
J.~C. Lashley, A.~C. Lawson, R.~J. McQueeney, and G.~H. Lander, cond-mat/0410634.
\bibitem{Richter01}
For a review and further references, see
M.~Richter,
in {\it Handbook of Magnetic Materials}, edited by
K.~H.~J.~Buschow (Elsevier, Amsterdam, 2001), Vol. 13, p. 87.
\bibitem{Ek93}
J.~van~Ek, P.~A.~Sterne, and A.~Gonis,
Phys. Rev. B {\bf 48}, 16280 (1993).
\bibitem{Soderlind97}
P. S\"oderlind, J.~M.~Wills, B.~Johansson, and O.~Eriksson,
Phys. Rev. B {\bf 55}, 1997 (1997).
\bibitem{Jones00}
M.~D.~Jones, J.~C.~Boettger, R.~C.~Albers, and D.~J.~Singh,
Phys. Rev. B {\bf 61}, 4644 (2000).
For the calculations of other plutonium phases with FLAPW-GGA method, see,
P.~S\"oderlind,
Adv. Phys. {\bf 47}, 959 (1998);
P. S\"oderlind and B.~Sadigh,
Phys. Rev. Lett. {\bf 92}, 185702 (2004);
G.~Robert, A.~Pasturel, and B.~Siberchicot,
J. Phys.: Condens. Matter {\bf 15}, 8377 (2003);
B.~Sadigh, P.~S\"oderlind, and W.~G.~Wolfer,
Phys. Rev. B {\bf 68}, 241101(R) (2003).
VASP code was used by
J.~Bouchet, R.~C.~Albers, M.~D.~Jones, and G.~Jomard,
Phys. Rev. Lett. {\bf 92}, 095503 (2004).
\bibitem{Soderlind94}
P.~S\"oderlind, O.~Eriksson, B.~Johansson, and J.~M.~Wills,
Phys. Rev. B {\bf 50}, 7291 (1994);
J.~M.~Wills and O.~Eriksson,
{\it ibid.} {\bf 45}, 13879 (1992).
\bibitem{Fernando00}
G.~W.~Fernando, E.~H.~Sevilla, and B.~R.~Cooper,
Phys. Rev. B {\bf 61}, 12562 (2000).
\bibitem{Nordstrom01}
L.~Nordstr\"om, J.~M.~Wills, P.~H.~Andersson, P.~S\"oderlind, and O.~Eriksson,
Phys. Rev. B {\bf 63}, 035103 (2001).
\bibitem{LDA+U}
For the review, see {\it Strong Coulomb Correlations in Electronic
Structure Calculations: Beyond the Local Density Approximation},
edited by V.~I.~Anisimov (Gordon and Breach Science Publishers, Amsterdam, 2000);
V.~I.~Anisimov, F.~Aryasetiawan, and A.~I.~Lichtenstein,
J. Phys.: Condens. Matter {\bf 9}, 767 (1997).
\bibitem{Savrasov00}
S.~Y.~Savrasov and G.~Kotliar,
Phys. Rev. Lett. {\bf 84}, 3670 (2000).
\bibitem{Lieser74}
{\it Plutonium -- A General Servey},
edited by K.~H.~Lieser (Verlag, Chemie, 1974).
\bibitem{SavrasovJ}
The set of Slater integrals $F^{(2)}$ = 10~eV, $F^{(4)}$ = 7~eV,
and $F^{(6)}$ = 5~eV used in Ref.~\onlinecite{Savrasov00} leads to Hund
exchange parameter $J_H$ = 0.85~eV according to the standard
formula $J_H$~=~(286$F^{(2)}$ + 195$F^{(4)}$ + 250$F^{(6)}$)/6435
for $f$ elements.
\bibitem{Bouchet00}
J.~Bouchet, B.~Siberchicot, F.~Jollet, and A.~Pasturel,
J.~Phys.:~Condens. Matter {\bf 12}, 1723 (2000).
\bibitem{Soderlind02}
P.~S\"oderlind, A.~L.~Landa, and B.~Sadigh,
Phys. Rev. B {\bf 66}, 205109 (2002).
\bibitem{Soderlind01}
P.~S\"oderlind,
Europhys. Lett. {\bf 55}, 525 (2001).
\bibitem{Wang01}
Y.~Wang and Y.~Sun,
J. Phys.: Condens. Matter {\bf 12}, L311 (2000).
\bibitem{Postnikov00}
A.~V.~Postnikov and V.~P.~Antropov,
Comp. Mat. Sci. {\bf 17}, 438 (2000).
\bibitem{Solovyev91}
I.~V.~Solovyev, A.~I.~Liechtenstein, V.~A.~Gubanov, V.~P.~Antropov, and O.~K.~Andersen,
Phys. Rev. B {\bf 43}, 14414 (1991);
M.~I.~Katsnelson, I.~V.~Solovyev, and A.~V.~Trefilov,
JETP Lett. {\bf 56}, 272 (1992).
\bibitem{Penicaud97}
M.~P\'enicaud,
J. Phys.: Condens. Matter {\bf 9}, 6341 (1997).
\bibitem{Singh94}
D.~J.~Singh,
{\it Planewaves, Pseudopotentials and the LAPW Method}
(Kluwer, Boston, 1994).
\bibitem{Kutepov03}
A.~L.~Kutepov and S.~G.~Kutepova,
J. Phys.: Condens. Matter {\bf 15}, 2607 (2003).
Compare with
X.~Wu and A.~K.~Ray,
cond-mat/0407676.
\bibitem{Kutepov04}
A.~L.~Kutepov and S.~G.~Kutepova,
J. Magn. Magn. Mater. {\bf 272-276}, e329 (2004).
\bibitem{Eriksson99}
O.~Eriksson, J.~D.~Becker, A.~V.~Balatsky, and J.~M.~Wills,
J. Alloys Comp. {\bf 287}, 1 (1999). See also 
B.~R. Cooper, O. Vogt, Q.-G. Sheng, and Y.-L. Lin, 
Philos. Mag. B {\bf 79}, 683 (1999). 
\bibitem{Petit01}
L.~Petit, A.~Svane, Z.~Szotek, P.~Strange, H.~Winter, and W.~M.~Temmerman,
J. Phys.: Condens. Matter {\bf 13}, 8697 (2001);
L.~Petit, A.~Svane, W.~M.~Temmerman, and Z.~Szotek,
Solid State Commun. {\bf 116}, 379 (2000).
\bibitem{Perdew81}
J.~P. Perdew and A. Zunger, Phys. Rev. B {\bf 23}, 5048 (1981).
\bibitem{Temmerman98}
W. M. Temmerman, A. Svane, Z. Szotek, and H. Winter, in {\it
Electronic Density Functional Theory: Recent Progress and New
Directions}, edited by J.~F. Dobson, G. Vignale, and M. P. Das
(Plenum, New York, 1998), p. 327.
\bibitem{SIC}
L.~Petit, A.~Svane, Z.~Szotek, and W.~M.~Temmerman,
Mol. Phys. Rep. {\bf 38}, 20 (2003).
\bibitem{Petit04}
L.~Petit, A.~Svane, Z.~Szotek, and W.~M.~Temmerman,
Mat. Res. Soc. Symp. Proc. {\bf 802}, DD6.7.1 (2004).
\bibitem{Petit02}
L.~Petit, A.~Svane, W.~M.~Temmerman, and Z.~Szotek,
J. Eur. Phys. B {\bf 25}, 139 (2002).
\bibitem{Petit03}
L.~Petit, A.~Svane, and W.~M.~Temmerman,
Science {\bf 301}, 498 (2003).
\bibitem{Niklasson03}
A.~M.~N.~Niklasson, J.~M.~Wills, M.~I.~Katsnelson, I.~A.~Abrikosov, O.~Eriksson, and B.~Johansson,
Phys. Rev. B {\bf 67}, 235105 (2003).
\bibitem{Wills04}
J.~M.~Wills, O.~Eriksson, A.~Delin, P.~H.~Andersson, J.~J.~Joyce, T.~Durakiewicz,
M.~T.~Butterfield, A.~J.~Arko, D.~P.~Moore, and L.~A.~Morales,
J. Elec. Spectrosc. Rel. Phen. {\bf 135}, 163 (2004).
\bibitem{KotliarVollhardt}
G.~Kotliar and D.~Vollhardt,
Phys. Today {\bf 57}, No. 3 (March), 53 (2004).
\bibitem{DMFT}
A.~Georges, G.~Kotliar, W.~Krauth, and M.~J.~Rozenberg,
Rev. Mod. Phys. {\bf 68}, 13 (1996).
\bibitem{condmat}
A.~Georges,
AIP Conf. Proc. {\bf 715}, 3 (2004);
G.~Kotliar and S.~Y.~Savrasov, in
{\it New Theoretical Approaches to Strongly Correlated Systems},
edited by A.~M. Tsvelik
(Kluwer Academic Publishers, The Netherlands, 2001), p.259.
\bibitem{Anis97}
V.~I. Anisimov, A.~I. Poteryaev, M.~A. Korotin, A.~O. Anokhin, and G. Kotliar,
J. Phys.: Condens. Matter {\bf 9}, 7359 (1997).
\bibitem{Held03}
K.~Held, I.~A.~Nekrasov, G.~Keller, V.~Eyert, N.~Bl\"umer, A.~K.~McMahan, R.~T.~Scalettar,
Th.~Pruschke, V.~I.~Anisimov, and D.~Vollhardt,
{\it Psi-k Newsletter} {\bf 56}, 65 (2003);
D.~Vollhardt, K.~Held, G.~Keller, R.~Bulla, Th.~Pruschke, I.~A.~Nekrasov, and
V.~I.~Anisimov,
cond-mat/0408266.
\bibitem{Savrasov04}
S.~Y.~Savrasov and G.~Kotliar,
Phys. Rev. B {\bf 69}, 245101 (2004);
G.~Kotliar and S.~Y.~Savrasov,
Int. J. Mod. Phys. B {\bf 17}, 5101 (2003).
\bibitem{Savrasov01}
S.~Y.~Savrasov, G.~Kotliar, and E.~Abrahams,
Nature {\bf 410}, 793 (2001). For the details of this calculation and comprehensive
review of the method, see Ref.~\onlinecite{Savrasov04}.
\bibitem{Benedict93}
B.~Johansson and M.~S.~S.~Brooks,
in {\it Handbook on the Physics and Chemistry of Rare Earths}, edited by
K.~A.~Gschneidner, Jr., L.~Eyring, G.~H.~Lander, and G.~R.~Choppin
(Elsevier, Amsterdam, 1993), Vol. 17, p. 149;
U.~Benedict and W.~B.~Holzapfel,
{\it ibid.}, Vol. 17, p. 245.
\bibitem{Solovyev98}
I.~V.~Solovyev, A.~I.~Liechtenstein, and K.~Terakura,
Phys. Rev. Lett. {\bf 80}, 5758 (1998).
\bibitem{LMTO}
O.~K.~Andersen, Phys. Rev. B {\bf 12}, 3060 (1975);
O.~Gunnarsson, O.~Jepsen, and O.~K.~Andersen,
Phys. Rev. B {\bf 27}, 7144 (1983).
\bibitem{Gunnarsson89}
O.~Gunnarsson, O.~K.~Andersen, O.~Jepsen, and J.~Zaanen,
Phys. Rev. B {\bf 39}, 1708 (1989).
\bibitem{Anisimov91a}
V.~I.~Anisimov and O.~Gunnarsson,
Phys. Rev. B {\bf 43}, 7570 (1991).
\bibitem{LandauSO}
L.~D.~Landau and E.~M.~Lifshitz, {\it Quantum Mechanics: Non-Relativistic
Theory} (Pergamon, Oxford, 1965).
\bibitem{DFT}
P.~Hohenberg and W.~Kohn,
Phys. Rev. {\bf 136}, B864 (1964);
W.~Kohn and L.~J.~Sham,
{\it ibid.} {\bf 140}, A1133 (1965).
\bibitem{vonBarth72}
U.~von~Barth and L.~Hedin,
J. Phys. C: Solid State Phys. {\bf 5}, 1629 (1972).
\bibitem{Sandratskii98}
For a comprehensive review of noncollinear magnetic calculations, see
L.~M.~Sandratskii,
Adv. Phys. {\bf 47}, 91 (1998).
\bibitem{Kubler88}
J.~K\"ubler, K.-H.~H\"ock, J.~Sticht, and A.~R.~Williams,
J. Phys. F: Met. Phys. {\bf 18}, 469 (1988).
\bibitem{Nordstrom96}
L.~Nordstr\"om and D.~J.~Singh,
Phys. Rev. Lett. {\bf 76}, 4420 (1996).
\bibitem{Young01}
D.~A.~Young,
{\it Phase diagram of the Elements}
(University of California Press, Berkley, 1991).
\bibitem{Dedericks84}
P.~H.~Dederichs, S.~Bl\"ugel, R.~Zeller, and H.~Akai,
Phys. Rev. Lett. {\bf 53}, 2512 (1984).
\bibitem{Desclaux84}
J.~P.~Desclaux and A.~J.~Freeman,
in {\it Handbook on the Physics and Chemistry of the Actinides}, edited by
A.~J.~Freeman and G.~H.~Lander (Elsevier, Amsterdam, 1984), Vol. 1.
\bibitem{Turchi99}
P.~E.~A.~Turchi, A.~Gonis, N.~Kioussis, D.~L.~Price, and B.~R.~Cooper,
in {\it Electron Correlations and Materials Properties}, edited by A.~Gonis,
N.~Kioussis, and M.~Ciftan (Kluwer Academic, New York, 1999), p. 531.
\bibitem{signnote}
Vectors of spin $\mathbf S$ and orbital $\mathbf L$ moments have
opposite directions so that total moments value is calculated via
$J=|L-S|$.
\bibitem{SolovyevUJ}
I.~Solovyev, N.~Hamada, and K.~Terakura, 
Phys. Rev. B {\bf 53}, 7158 (1996).
\bibitem{PuNStr}
F.~Anselin,
J. Nucl. Mater. {\bf 10}, 331 (1963).
\bibitem{Boeuf84}
A.~Boeuf, R.~Cacciuffo, J.~M.~Fournier, L.~Manes, J.~Rebizant, E.~Roudaut, and
F.~Rustichelli,
Solid State Commun. {\bf 52}, 451 (1984). See also
P.~Burlet, J.~M.~Fournier, L.~Manes, J.~Rebizant, and F.~Rustichelli, Proc. of Ili\'emes
Joun\'ees des Actinides, edited by G.~Bombieri (Venice, Italy, 1982).
\bibitem{Martin76}
D.~J.~Martin, R.~D.~A.~Hall, J.~A.~Lee, M.~J.~Mortimer, and P.~W.~Sutcliffe,
Harwell AERE Rep. {\bf 76}, 12599 (1976).
\bibitem{Raphael69}
G.~Rapha\"el and C. H. de Novion,
Solid State Commun. {\bf 7}, 791 (1969).
\bibitem{Brooks84}
M.~S.~S.~Brooks,
J. Phys. F: Met. Phys. {\bf 14}, 857 (1984).
\bibitem{Sarrao02}
J.~L.~Sarrao, L.~A.~Morales, J.~D.~Thompson, B.~L.~Scott,
G.~R.~Stewart, F.~Wastin, J.~Rebizant, P.~Boulet, E.~Colineau, and G.~H.~Lander,
Nature {\bf 420}, 297 (2002).
\bibitem{Opahle04}
I. Opahle, S. Elgazzar, K. Koepernik, and P.~M. Oppeneer,
Phys. Rev. B {\bf 70}, 104504 (2004).
\bibitem{115a}
H. Hegger, C. Petrovic, E.~G. Moshopoulou, M.~F. Hundley, J.~L. Sarrao,
Z. Fisk, and J.~D. Thompson,
Phys. Rev. Lett. {\bf 84}, 4986 (2000);
C. Petrovic, P.~G. Pagliuso, M.~F. Hundley, R. Movshovich, J.~L. Sarrao, J.~D. Thompson,
Z. Fisk, and P. Monthoux,
J. Phys.: Condens. Matter {\bf 13}, L337 (2001);
C. Petrovic, R. Movshovich, M. Jaime, P.~G. Pagliuso, M.~F. Hundley, J.~L. Sarrao,
Z. Fisk, and J.~D. Thompson,
Europhys. Lett. {\bf 53}, 354 (2001).
\bibitem{115b}
M.~F. Hundley, A. Malinowski, P.~G. Pagliuso, J.~L. Sarrao, and J.~D. Thompson,
Phys. Rev. B {\bf 70}, 035113 (2004).
\bibitem{Sarrao03}
J.~L.~Sarrao, J.~D.~Thompson, N.~O.~Moreno, L.~A.~Morales, F.~Wastin, J.~Rebizant,
P.~Boulet, E.~Colineau, and G.~H.~Lander,
J. Phys.: Condens. Matter {\bf 15}, S2275 (2003).
\bibitem{Thompson04}
J.~D.~Thompson, J.~L.~Sarrao, L.~A.~Morales, F.~Wastin, and P.~Boulet,
Physica C {\bf 412-414}, 10 (2004).
\bibitem{Griveau04}
J.-C.~Griveau, C.~Pfleiderer, P.~Boulet, J.~Rebizant, and F.~Wastin,
J. Magn. Magn. Mater. {\bf 272-276}, 154 (2004).
\bibitem{Bang04}
Y.~Bang, A.~V.~Balatsky, F.~Wastin, and J.~D.~Thompson,
Phys. Rev. B {\bf 70}, 104512 (2004).
\bibitem{Bauer04}
E.~D. Bauer, J.~D. Thompson, J.~L. Sarrao, L.~A. Morales, F. Wastin, J. Rebizant,
J.~C. Griveau, P. Javorsky, P. Boulet, E. Colineau, G.~H. Lander, and G.~R. Stewart,
Phys. Rev. Lett. {\bf 93}, 147005 (2004).
\bibitem{Tanaka04}
K.~Tanaka, H.~Ikeda, and K.~Yamada,
cond-mat/0403069.
\bibitem{Wastin03}
F.~Wastin, P.~Boulet, J.~Rebizant, E.~Colineau, and G.~H.~Lander,
J. Phys.: Condens. Matter {\bf 15}, S2279 (2003).
\bibitem{Opahle03}
I.~Opahle and P.~M.~Oppeneer,
Phys. Rev. Lett. {\bf 90}, 157001 (2003).
\bibitem{Szajek03}
A.~Szajek and J.~A.~Morkowski,
J. Phys.: Condens. Matter {\bf 15}, L155 (2003).
\bibitem{Maehira03}
T.~Maehira, T.~Hotta, K.~Ueda, and A.~Hasegawa,
Phys. Rev. Lett. {\bf 90}, 207007 (2003); cond-mat/0212033.
\bibitem{Hotta03}
T.~Hotta and K. Ueda,
Phys. Rev. B {\bf 67}, 104518 (2003).
\bibitem{Soderlind04a}
P.~S\"oderlind,
Phys. Rev. B {\bf 70}, 094515 (2004).
\bibitem{Harvey73}
A.~R.~Harvey, M.~B.~Brodsky, and W.~J.~Nellis,
Phys. Rev. B {\bf 7}, 4137 (1973).
\bibitem{Stewart84}
G.~R.~Stewart, B.~Andraka, and R.~G.~Haire,
J. Alloys Comp. {\bf 213}, 111 (1994).
\bibitem{Boulet03}
P.~Boulet, F.~Wastin, E.~Colineau, J.~C.~Griveau, and J.~Rebizant,
J. Phys.: Condens. Matter {\bf 15}, S2305 (2003).
\bibitem{Zachariasen49}
W.~H.~Zachariasen,
Acta Crystallogr. {\bf 2}, 94 (1949);
C.~C.~Land, K.~A.~Johnson, and F.~H.~Ellinger,
J. Nucl. Mater. {\bf 15}, 23 (1965).
\bibitem{Olsen60}
S.~E.~Olsen,
J. Appl. Phys. {\bf 31}, 340 (1960).
\bibitem{Fournier90}
J.~M.~Fournier, E.~Pleska, J.~Chiapusio, J.~Rossat-Mignod, J.~Rebizant, J.~C.~Spirlet,
and O.~Vogt,
Physica B {\bf 163}, 493 (1990).
\bibitem{Stewart91}
G.~R.~Stewart, R.~G.~Haire, J.~C.~Spirlet, and J.~Rebizant,
J. Alloys Comp. {\bf 177}, 167 (1991).
\bibitem{Lander88}
G.~H.~Lander, M.~Wulff, J.~Rebizant, J.~C.~Spirlet, P.~J.~Brown, and O.~Vogt,
J. Appl. Phys. {\bf 63}, 3601 (1988).
\bibitem{Wachter91}
P.~Wachter, F.~Marabelli, and B.~Bucher,
Phys. Rev. B {\bf 43}, 11136 (1991).
\bibitem{Mendik93}
M.~Mendik, P.~Wachter, J.~C.~Spirlet, and J.~Rebizant,
Physica B {\bf 186-188}, 678 (1993).
\bibitem{Wachter03}
P.~Wachter,
Solid State Commun. {\bf 127}, 599 (2003).
\bibitem{Burlet89}
P.~Burlet, S.~Quezel, J.~Rossat-Mignod, J.~C.~Spirlet, J.~Rebizant, and O.~Vogt,
Physica B {\bf 159}, 129 (1989);
G.~H.~Lander, J.~Rebizant, J.~C.~Spirlet, A.~Delapalme, P.~J.~Brown, O.~Vogt, and
K.~Mattenberger,
Physica B {\bf 146}, 341 (1987).
\bibitem{Olsen92}
C.~E.~Olsen, A.~L.~Comstock, and Th.~A.~Sandenaw,
J. Nucl. Mater. {\bf 195}, 312 (1992).
\bibitem{Abraham96}
C.~Abraham, U.~Benedict, and J.~C.~Spirlet,
Physica B {\bf 222}, 52 (1996).
\bibitem{Ichas01}
V.~Ichas, J.~C.~Griveau, J.~Rebizant, and J.~C.~Spirlet,
Phys. Rev. B {\bf 63}, 045109 (2001).
\bibitem{Brooks87}
M.~S.~S.~Brooks,
J. Magn. Magn. Mater. {\bf 63-64}, 649 (1987).
\bibitem{Hasegawa92}
A.~Hasegawa and H.~Yamagami,
J. Magn. Magn. Mater. {\bf 104-107}, 65 (1992).
\bibitem{Oppeneer00}
P.~M.~Oppeneer, T.~Kraft, and M.~S.~S.~Brooks,
Phys. Rev. B {\bf 61}, 12825 (2000).
\bibitem{Sanchez87}
J.~P.~Sanchez, J.~C.~Spirlet, J.~Rebizant, and O.~Vogt,
J. Magn. Magn. Mater. {\bf 63-64}, 139 (1987).
\bibitem{Cooper83}
B.~R.~Cooper, P.~Thayamballi, J.~C.~Spirlet, W.~M\"uller, and O.~Vogt,
Phys. Rev. Lett. {\bf 51}, 2418 (1983);
G.~H.~Lander, A.~Delapalme, P.~J.~Brown, J.~C.~Spirlet, J.~Rebizant,
and O.~Vogt,
J. Appl. Phys. {\bf 57}, 3748 (1985).
\bibitem{Lander84}
G.~H.~Lander, A.~Delapalme, P.~J.~Brown, J.~C.~Spirlet, J.~Rebizant,
and O.~Vogt,
Phys. Rev. Lett. {\bf 53}, 2262 (1984).
\bibitem{Burlet84}
P.~Burlet, S.~Quezel, J.~Rossat-Mignod, J.~C.~Spirlet, J.~Rebizant, W.~M\"uller, and O.~Vogt,
Phys. Rev. B {\bf 30}, 6660 (1984);
P.~Burlet, J.~Rossat-Mignod, G.~H.~Lander, J.~C.~Spirlet, J.~Rebizant, and O.~Vogt,
{\it ibid.} {\bf 36}, 5306 (1987).
\bibitem{Lander86}
G.~H.~Lander, W.~G.~Stirling, J.~Rossat-Mignod, J.~C.~Spirlet, J.~Rebizant, and O.~Vogt,
Physica B {\bf 136}, 409 (1986).
\bibitem{Spirletunpub}
J.~C.~Spirlet, J. Rebizant, and O. Vogt,
in Treizi\`eme Journ\`ee des Actinides, Elat, Israel, 1983 (unpublished).
\bibitem{Hu87}
G.~J.~Hu, N.~Kioussis, B.~R.~Cooper, and A.~Banerjea,
J. Appl. Phys. {\bf 61}, 3385 (1987).
\bibitem{Blaise85}
A.~Blaise, J.~M.~Collard, J.~M.~Fournier, J.~Rebizant, J.~C.~Spirlet, and O.~Vogt,
Physica B {\bf 130}, 99 (1985).
\bibitem{Gouder00}
T.~Gouder, F.~Wastin, J.~Rebizant, and L.~Havela,
Phys. Rev. Lett. {\bf 84}, 3378 (2000).
\bibitem{Durakiewicz04}
T. Durakiewicz, J.~J. Joyce, G.~H. Lander, C.~G. Olson, M.~T. Butterfield,
E. Guziewicz, A.~J. Arko, L. Morales, J. Rebizant, K. Mattenberger, and O. Vogt,
Phys. Rev. B {\bf 70}, 205103 (2004).
\end{thebibliography}
\end {document}